\documentclass[12pt]{article}
\usepackage{tikz}
\usetikzlibrary{positioning,shapes.geometric,fit}
\usepackage{rotating}
\usepackage{adjustbox,lipsum}
\renewcommand{\baselinestretch}{1.5}

\usepackage{amsmath, amsfonts}
\usepackage{bm, bbm}
\usepackage{graphics, graphicx, color, epsf}
\usepackage{natbib}
\usepackage{xr, zref, hyperref}
\usepackage{algorithm, algorithmic} 

\usepackage[hang,flushmargin]{footmisc} 

\usepackage{etoolbox, siunitx} 
\robustify\bfseries

\newcommand{\E}{\mathrm{E}}
\newcommand{\var}{\mathrm{var}}

 \newcommand{\Appendix}
 {
 \def\thesection{\Alph{section}}
 \def\thesubsection{\Alph{section}.\arabic{subsection}}
 \def\theequation{\Alph{section}.\arabic{equation}}
 \def\thefigure{\Alph{section}.\arabic{figure}}
 \def\thealg{\Alph{section}.\arabic{alg}}
 }

\makeatletter
\newcommand*{\addFileDependency}[1]{
  \typeout{(#1)}
  \@addtofilelist{#1}
  \IfFileExists{#1}{}{\typeout{No file #1.}}
}
\makeatother

\begin{document}

\def\spacingset#1{\renewcommand{\baselinestretch}%
{#1}\small\normalsize} \spacingset{1}

\vskip 2mm

{
  \title{\bf Interpolating Population Distributions using Public-use Data: An Application to Income Segregation using American Community Survey Data}
  \author{Matthew Simpson\thanks{This research was partially supported by the U.S. National Science Foundation (NSF) and the U.S. Census Bureau under NSF grant SES-1132031, funded through the NSF-Census Research Network (NCRN) program, and under NSF grant SES-1853096. This article is released to inform interested parties of research and to encourage discussion.  The views expressed on statistical issues are those of the authors and not those of the NSF or the U.S. Census Bureau.}\thanks{The computation for this work was performed on the high performance computing infrastructure provided by Research Computing Support Services and in part by the National Science Foundation under grant number CNS-1429294 at the University of Missouri, Columbia MO.}\thanks{The authors thank Noel Cressie for helpful discussion on earlier versions of this manuscript.}\hspace{.2cm}\\
    SAS Institute\\
    (to whom correspondence should be addressed)\\
    Matt.Simpson@sas.com\\\\
     Scott H. Holan\\
     Department of Statistics, University of Missouri,\\
     U.S. Census Bureau\\\\
     Christopher K. Wikle\\
       Department of Statistics, University of Missouri\\\\
       and\\\\
       Jonathan R. Bradley\\
       Department of Statistics, Florida State University}
     \maketitle
}

\newpage
\begin{abstract}
Income segregation measures the extent to which households choose to live near other households with similar incomes. Sociologists theorize that income segregation can exacerbate the impacts of income inequality, and have developed indices to measure it at the metro area level, including the information theory index introduced in \citet{reardon2011income}, and the divergence index presented in \citet{roberto2015divergence}. To study their differences, we construct both indices using recent American Community Survey (ACS) estimates of features of the income distribution. Since the elimination of the decennial census long form, methods of computing these estimates must be updated to use ACS estimates and account for survey error. We propose a model-based method to interpolate estimates of features of the income distribution that accounts for this error. This method improves on previous approaches by allowing for the use of more types of estimates, and by providing uncertainty quantification. We apply this method to estimate U.S. census tract-level income distributions using ACS tabulations, and in turn use these to construct both income segregation indices. We find major differences between the two indices in the relative ranking of metro areas, as well as differences in how both indices correlate with the Gini index.
\end{abstract}

\noindent%
\vfill

\newpage
\spacingset{1.45} 

\section{INTRODUCTION}\label{sec:intro}
Sociologists theorize that peer or neighborhood effects of income segregation can exacerbate the impacts of income inequality \citep[][and references therein]{reardon2011income} -- households are segregated by income to the extent that households with similar incomes live near each other. To study this, \citet{reardon2011measures} and \citet{reardon2011income} develop the rank-order information theory index of income segregation in metro areas, which essentially compares the metro area income distribution to the census tract-level income distributions for each tract in the metro area. They then fit various regression models to the index using decennial census data. More recently, \citet{roberto2015divergence} notes that the information theory index can give results that conflict with our intuitions about the meaning of income segregation, and suggests an alternative index based on the Kullback–Leibler (KL) divergence, called the divergence index.

Both of these indices require as inputs tract-level income distributions for each census tract within a given metro area. The \citet{reardon2011income} analysis relied on decennial census data, which provides detailed distributional information at the tract level and has no associated survey error. Since the elimination of the decennial census long form, this information is no longer available, and instead data users must rely on American Community Survey (ACS) estimates with associated standard errors. The ACS provides fairly detailed information about tract-level income distributions in the form of \emph{bin estimates}; i.e., estimates of the proportion or number of households in a given census tract with an income in a small number of income bins. For example, Table~\ref{tab:acs} in Appendix~\ref{app:explore} of the Supplementary Materials contains 2015 ACS 5-year period bin estimates for several census tracts in Boone County, MO.

Our goal is to use these and other estimates of features of the tract-level income distributions to estimate each tract-level income distribution. Then, in turn, we use these distributions to construct both income segregation indices and reproduce a portion of the \citet{reardon2011income} analysis using both indices and more recent ACS data. Specifically, we assess the degree to which the Gini index, a measure of income inequality, predicts income segregation as measured by both indices at the household level, and at the household-race level. The regressions we fit to study this relationship must account for the uncertainty in the ACS estimates we use as covariates, as well as in the estimated indices we use as responses. We fit error-in-variables (EIV) regressions to account for this uncertainty. We find that the two indices substantially disagree about the relative ranking of U.S. metro areas in terms of income segregation. Additionally, the correlation between income inequality as measured by the Gini index and income segregation can depend on which index is used, though this difference is less stark.

Our methodological contribution is to construct tract-level income distributions using only ACS estimates of features of those distributions. Many authors use a method called the ``Pareto-linear procedure'' (PRLN) to construct these distributions using bin estimates, typically as an intermediate step to obtain an estimate of the Gini index, e.g., \citet{jargowsky1996take,nielsen1997kuznets,hipp2007block,hipp2007income,moller2009changing,hipp2013extrapolative,braithwaite2015sexual}, among others. PRLN assumes that income is uniformly distributed within bins that include or are below the median, and Pareto distributed in bins above the median, with some exceptions to handle special cases. The methodology is well-established, and is effective for income distributions \citep{miller1966income,aigner1970estimation,kakwani1976efficient,spiers1977prln,henson1980money,welniak1988calculating}. 

However, PRLN suffers from several limitations, especially with respect to our problem. First, PRLN does not quantify uncertainty about the income distribution -- it only provides a point estimate. Thus, confidence intervals and standard errors are not available for estimates of the Gini index or segregation indices based on PRLN. Second, PRLN is only able to use bin estimates. The ACS provides many other estimates of features of the income distribution including quantiles, income shares, and the Gini index. Taking these into account should result in more accurate estimates of the income distribution of interest. Third, PRLN does not take into account the standard error associated with the estimates that it does use. This is understandable given that PRLN was designed to be used with decennial census data. However, if data users ignore the standard errors in ACS data, their analyses will understate uncertainty (to the extent they quantify uncertainty at all) and potentially be biased.

We solve these three issues with PRLN by taking a latent density estimation approach based on PRLN, which we call latent PRLN (L-PRLN). This approach is able to take into account multiple diverse types of estimates associated with a given distribution, and naturally accounts for the inherent uncertainty associated with the estimates used by the model. These estimates are estimates of functionals of the latent tract-level income distributions, so our model borrows elements from functional data analysis (FDA) -- see e.g., \citet{ramsay2006functional}, \citet{ferraty2006nonparametric}, and \citet{kokoszka2017introduction} for overviews. However, our case differs from the usual FDA case because the latent functions we are attempting to estimate are probability distribution functions (PDFs), or equivalently any function that uniquely determines the latent probability distribution such as a cumulative distribution function (CDF) or quantile function. This puts constraints on the latent function that are not typical for FDA, and necessarily implies a different modeling strategy. There are several small area estimation (SAE) approaches concerned with estimation of income and other related quantities (such as poverty or per capita household expenditures etc.) Nevertheless, these are typically unit-level models that directly use income (or other proxy variable) \citep[e.g., see][among others]{battese1988error,elbers2003micro,marchetti2012non,molina2010small,tarozzi2009using,tzavidis2008m} or area-level models \citep[e.g., see][]{fay1979estimates} that use a direct estimator of income as the model inputs. In contrast, our model uses features of the income distribution as the model inputs rather than income (or another proxy variable) directly. Specifically, we can not fit the SAE models previously listed based on the information that encompasses our model inputs.

Similarly, our approach is also related to the literature on density estimation. The most popular approach is kernel density estimation \citep[e.g.][]{scott2015multivariate}, but this approach does not directly apply to our setting since we do not have observations drawn from the distribution of interest. Another approach is log splines \citep{kooperberg1992logspline, stone1994use}, which is subject to the same criticism for our problem. In essence, however, our model is fundamentally inspired by PRLN and can be motivated from that perspective. The choice of PRLN as a starting point for our model was in part based on computational convenience as well as its semiparametric specification. Nevertheless, other parametric alternatives could also be considered, e.g., see \citet{singh1976function} and \citet{Dagum1977} among others. However, purely parametric approaches may be less flexible accross a broad array of applications.

The remainder of the paper is organized as follows. In Section~\ref{sec:acs.model} we begin by describing the ACS and available estimates of features of the income distribution, then in Section~\ref{sec:PRLN} we describe PRLN, and use it to motivate L-PRLN in Section~\ref{sec:model}. In Section~\ref{sec:proof} we compare L-PRLN and PRLN in a pair of tests. First, in Section~\ref{sec:sim} we conduct a simulation study where we repeatedly sample from a fixed synthetic population and fit both models to each sample. Second, in Section~\ref{sec:acs.test} we fit both models to ACS data and compare model-based estimates to held-out direct estimates of various features of the income distributions. Next, in Section~\ref{sec:index}, we return to the income segregation index problem. Here we describe both indices, estimate both of them using ACS data, then use both in a partial reproduction of the analysis of \citet{reardon2011income} using more recent ACS data. Finally, in Section~\ref{sec:discuss}, we discuss our results and conclude. Supplementary material includes several appendices referenced in the paper.

\section{AMERICAN COMMUNITY SURVEY AND MODEL MOTIVATION}\label{sec:acs.model}
The U.S. Census Bureau administers the ACS to produce a variety of annually released data products used by public and private institutions. There are two main types of data products. First, ACS estimates of various quantities are tabulated and published for several geographies, including census tracts, counties, states, and national. Second, raw data files in the form of Public-Use Microdata Samples (PUMS) are released to the public. The PUMS are organized into PUMAs, and they contain a weighted sample of households and of residents living in each PUMA; more detailed location information about these residents and households is not available due to disclosure limitations. Each PUMA is designed to contain around $100,000$ people, and census tracts are nested within PUMAs.

The PUMS sample in a given PUMA for a given period is a subset of the full ACS sample for that same area and period, and the sample weights in the PUMS are not the same as the weights used to construct the ACS estimates \citep{pums2015}. Both the ACS estimates and PUMS are currently published based on one and five years of the survey, known as 1-year and 5-year period estimates and PUMS, respectively. Though areal units with less than 65,000 people only have published 5-year period estimates, in previous years areal units with at least 20,000 people also had published 3-year period estimates \citep{acs2014design}.

At the PUMA level, the PUMS provides detailed distributional information about a wide variety of variables measured on households and individuals. At the tract level, however, only a set of specific estimates are available. Many variables only have basic summary statistics published, such as means. Some variables, such as household income or age of householder, have more detailed information available, though not necessarily the information a data user is interested in. In 2015 the ACS published the following 5-year tract-level income distribution period estimates: mean income, median income, Gini index of income, the 20th, 40th, 60th, 80th, and 95th percentiles of income, income shares of each quintile and the top 5\% of the income distribution, and the proportion of households with incomes in 12 income bins defined by the following breaks: $\$5,000$, $\$10,000$, $\$15,000$, $\$20,000$, $\$25,000$, $\$35,000$, $\$50,000$, $\$75,000$, $\$100,000$, $\$150,000$, and $\$200,000$ \citep{acs2015quintiles,acs2015shares,acs2015gini,acs2015income,acs2015financial}. Each tract-level estimate also has a corresponding margin of error (MOE) so that $\mathrm{estimate} \pm \mathrm{MOE}$ determines a 90\% confidence interval, and $\mathrm{MOE}/1.645$ is the standard error of the estimate.

\subsection{The Pareto-linear procedure}\label{sec:PRLN}
The fundamental problem is to estimate a density $\pi$ using estimates of various features of that density. PRLN does this by only using the bin estimates. Let $k=1,2,\dots,K$ index bins, and let $\kappa_1 = 0 < \kappa_2 < \cdots < \kappa_K < \kappa_{K+1} = \infty$, denote the bin boundaries, which we will refer to as knots. Then the PRLN density is given by
\begin{align}
\pi(x) = \sum_{k=1}^Kp_kf_k(x). && \mbox{ (PRLN density) } \label{eq:density.model}
\end{align}
where $p_k$ is the probability associated with bin $k$, and $f_k$ is the probability density within bin $k$, with support $(\kappa_k, \kappa_{k+1}]$, except in the uppermost bin where $f_K$ has the support $(\kappa_K, \infty)$. Let $k^*$ denote the index of largest knot below the median according to the bin estimates. Then the PRLN density defines the $f_k$s via
\begin{align}
  f_k(x) = &\  \frac{1}{\kappa_{k+1} - \kappa_k} \times \mathbbm{1}(\kappa_k < x \leq \kappa_{k+1}) && \mbox{ if $k \leq k^*$ },\nonumber\\
  = &\  \frac{\alpha_k\kappa_k^{\alpha_k}x^{-\alpha_k - 1}}{1 - \left(\frac{\kappa_k}{\kappa_{k+1}}\right)^{\alpha_k}} \times \mathbbm{1}(\kappa_k < x \leq \kappa_{k+1}) && \mbox{ if $k^* < k < K$ },\nonumber\\
  = &\  \alpha_k\kappa_k^{\alpha_K}x^{-\alpha_k - 1} \times \mathbbm{1}(\kappa_K < x) && \mbox{ if $k = K$ }\label{eq:prln.density}.
\end{align}
The unknown parameters of the model, which need to be estimated, are the knot probabilities, $\bm{p} = (p_1, p_2, \dots, p_K)$, as well as the Pareto parameters, $\bm{\alpha} = (\alpha_{k^* + 1}, \alpha_{k^* + 2}, \dots, \alpha_{K})$.

PRLN estimates the $p_k$s with the associated bin estimates, which we will denote by $b_k$ for $k=1,2,\dots,K$. Then PRLN estimates the Pareto parameters as follows. Let $B_k = \sum_{i=k}^Kb_i$ for $k=1,2,\dots,K$. The initial PRLN estimate for $\alpha_k$ is given by
\begin{align*}
\widehat{\alpha}_k = \log(B_k/B_{k-1}) / \log(\kappa_k / \kappa_{k-1}).
\end{align*}
If $\widehat{\alpha}_k \leq 1$, then in the truncated Pareto bins, PRLN reverts to a uniform distribution. In the uppermost bin, which is untruncated Pareto distributed, PRLN instead tries to use $\widehat{\alpha}_{K-1}$, i.e. the $\widehat{\alpha}$ from the bin just below it, as long as that bin was truncated Pareto distributed. If $\widehat{\alpha}_{K-1} \leq 1$, then it tries to use $\widehat{\alpha}_{K-2}$, and so on, until it reaches the last Pareto distributed bin. If it runs out of Pareto bins in this manner, then PRLN assumes that the uppermost bin is a point mass at the lower bound.

The PRLN density a judicious choice because there is not much information about the income distribution between the boundaries of the bins defining the bin estimates. This makes it difficult to estimate a large number of $p_k$s, or a larger number of parameters associated with the $f_k$s. The chosen knots help to minimize the number of $p_k$s as much as possible, and by assuming uniform distributions within the lower bins, PRLN further reduces the number of parameters to estimate. Additionally, since income distributions are known to have approximately Pareto right tails the Pareto bins are likely to fit well.

\subsection{L-PRLN: A semiparametric latent density model}\label{sec:model}
Despite its effectiveness, PRLN suffers from three major flaws for our purposes. It cannot quantify uncertainty and only provides point estimates, it cannot take into account all available estimates of features of the income distribution, and it does not take into account the standard error associated with the ACS estimates. Our key innovation to solve these problems is to treat the density as latent, and the published estimates as estimations of functionals of that density with some associated error. 

Let $u=1,2,\dots,U$ index the available published estimates, e.g. from the ACS, let $q_u$ denote the estimate and $S_u$ its standard error, and let $Q_u(\cdot)$ denote the functional that takes a probability distribution and returns the value of the estimand for that distribution. For example, if $q_u$ is an estimate of the mean, $Q_u(\pi) = \E_{\pi}[X]$. Typically a central limit theorem applies for the estimates, so we assume
\begin{align}
q_u|\pi, S_u \stackrel{ind}{\sim}\mathrm{N}(Q_u(\pi), S_u^2) && \mbox{ (data model) } \label{eq:data.model}
\end{align}
for $u=1,2,\dots,U$. The estimate errors are correlated, but these correlations are not available in the ACS, and in general are rarely publicly available. When they are available, \eqref{eq:data.model} can be modified appropriately to take into account the full error covariance matrix.

Next, we need a model for $\pi$. In theory, the class of densities used by log spline density estimation \citep{stone1994use} or kernel density estimation \citep{scott2015multivariate} could be used here, but a fundamental constraint is that we need to be able to compute $Q_u(\pi)$ quickly for many different $Q_u$s, including, for example, the mean of the density. Instead we use the PRLN density defined in \eqref{eq:density.model} and \eqref{eq:prln.density}. As long as $\alpha_K > 1$, then the CDF, quantile function, mean, income shares, and Gini coefficient of the PRLN density are all available in closed form. If $\alpha_K > 2$ then additionally the variance is available in closed form. See Appendix~\ref{app:functionals} for formulas for each of these functionals.

By treating the PRLN density as latent, we are able to solve all three limitations of PRLN. We easily take into account the standard errors and propagate that uncertainty into our estimates of the latent population and any distributional features of interest. Additionally, we are able to take into account a much wider variety of available estimates of features of the income distribution.

\subsection{Estimation and interpolation}\label{sec:post.pred}
To construct estimates of any feature of the distribution of interest, including interpolating between the end points of the bins, we will use the Bayesian posterior predictive distribution for the latent population in the area of interest. This allows us to construct a posterior distribution for any distributional feature of interest, so long as it can be easily computed for a finite population and we can easily simulate from $\pi$ conditional on its parameters. Additionally, it allows us to partially take into account the fact that the latent population is finite. We can even relax the finite population requirement so long as the distributional feature can be easily computed as a function of the model parameters.

We first must sample from the posterior of $\bm{\theta}$ to be able to sample from the posterior predictive distribution. We do this using the No-U-Turn Sampler \citep[NUTS;][]{hoffman2014nuts}, a variant of Hamiltonian Monte Carlo \citep[HMC;][]{neal2011hmc}. One reason for this choice is that conditional conjugacy in a Gibbs sampler is hopeless due to the form of the $Q_u$s. Additionally, NUTS tends to be more robust and efficient than other MCMC options even when conjugacy relationships are available \citep{betancourt2015hamiltonian}. We use the software package \verb0Stan0 \citep{gelman2015stan,rstan2016} to perform NUTS. NUTS requires the log-posterior and thus log-likelihood be available in closed form, up to an additive constant. The log-likelihood is implied by equation \eqref{eq:data.model} with the $Q_u$s defined in Appendix~\ref{app:functionals}.

To construct the posterior predictive distribution of the latent population, let $N$ denote an estimate of the population of the area of interest, e.g. from the ACS. Let $i=1,2\dots,N$ index the latent population, let $Y_i$ denote the $i$th latent income, and let $\bm{\theta}$ denote the full vector of unknown parameters. Then for each posterior sample $\bm{\theta}^{(m)}$, $m=1,2,\dots,M$ we generate the latent population via
\begin{align}
Y_i^{(m)} |\bm{\theta}^{(m)} \stackrel{iid}{\sim} \pi_{\bm{\theta}^{(m)}} && \mbox{ (posterior predictive distribution) }\label{eq:post.pred}
\end{align}
for $i=1,2,\dots,N$ and $m=1,2,\dots,M$. This is easily performed in a two step process. First, generate the bin the observation belongs to using $(p_1^{(m)}, \dots, p_K^{(m)})$ where $p_k$ denotes the probability of bin $k$. Then conditional on bin $k$ being chosen, $Y_i^{(m)}$ is generated from the density within that bin, $f_k$, conditional on $\bm{\theta}$, or more precisely the elements of $\bm{\theta}$ that determine $f_k$. Then the posterior distribution of any feature of the latent distribution of income can be obtained as a function of $\bm{Y}^{(m)} = (Y_1^{(m)}, Y_2^{(m)}, \dots, Y_N^{(m)})$ for $m=1,2,\dots,M$.

In principle, the standard error of $N$ can be taken into account by treating the true size of the population as an unknown, denoted by $\eta$, with estimate $N$ and standard error $H$. Then for each draw from the MCMC sampler, a new value of $\eta$ can be drawn via
\begin{align*}
\eta^{(m)} \stackrel{iid}{\sim}\mathrm{N}(N, H^2).
\end{align*}
Subsequently, $Y_i^{(m)}$ can be drawn via \eqref{eq:post.pred} for $i=1,2,\dots,\eta^{(m)}$. We do not use this approach here and, instead, treat the tract-level population estimates as the truth since it is unlikely to have a major impact on the results, but in cases where the population estimates are near zero and their standard errors are large, it may be worthwhile.

\subsection{Inverted quantile estimates}\label{sec:inverted.quantile}
To be able to use gradient based estimation methods such as NUTS, we use the delta method to ``invert'' the quantile data model. Suppose $q$ is an estimate of the $\tau$th quantile, $\Pi^{-1}(\tau)$, with standard error $S$. We originally assumed that $q \sim \mathrm{N}(\Pi^{-1}(\tau), S^2)$. Using the delta method we obtain \eqref{eq:invquant} as the data model for the corresponding \emph{inverted quantile estimate}, $\tau$,
\begin{align}
\tau | \pi, q, S \sim \mathrm{N}\left(\Pi(q), \left[\frac{S}{\pi(q)}\right]^2\right). \label{eq:invquant}
\end{align}
Since $\pi$ depends on several unknown parameters, NUTS is more difficult because it creates hard to eliminate divergences \citep[see e.g.][]{betancourt2015hamiltonian}. So we plug in an estimate of $\pi(q)$ using a modification of the original PRLN. See Section~\ref{sec:PRLN} for a description of PRLN's estimation process. We modify PRLN in two ways. First, our initial estimate of $\alpha_k$ is
\begin{align*}
\widehat{\alpha}_k = \left.\log\left(\frac{B_k + 0.0001}{B_{k-1} + 0.0001}\right) \right/ \log(\kappa_k / \kappa_{k-1}) 
\end{align*}
to prevent bins estimates of zero from causing problems. Second, instead of using a point mass as a last resort in the uppermost bin, we instead use $\widehat{\alpha}_K = 1.0001$. This is more realistic, and should result in a more accurate standard error, e.g., for $\tau = 0.95$. Note that for quantiles which are in bins that are uniform distributed, our plug-in estimate is $\widehat{\pi}(q) = b_{k^*} / (\kappa_{k^* + 1} - \kappa_{k^*})$ where $k^*$ is the index of the closest knot from below to $q$, and $b_{k^*}$ is the corresponding bin estimate.

\subsection{Priors}\label{subsec:priors}
To complete the model, we need to choose priors for the $p_k$s and the $\alpha_k$s. An extremely ``uninformative'' prior for $\bm{p}$ can cause problems for MCMC, so we opt for a weakly informative prior. Note also that the bins are not designed so that we would expect them to be equally probable {\it a priori}. Thus, we center $\bm{p}$ on the ACS 5-year period bin estimates for the entire United States, from the same year as the tract-level estimates, using a Dirichlet prior. Let $\bm{g}$ denote the country-level estimates, and let $t$ denote a scale hyperparameter, then we assume
\begin{align*}
\bm{p} \sim \mathrm{Dirichlet}(\bm{g} / t).
\end{align*}
The value of $t$ encodes the level of prior certainty that $\bm{g}$ is the true value of $\bm{p}$. A value of $t\geq 1$ is ideal since we do not necessarily expect $\bm{g}$ to be close to $\bm{p}$ with a high degree of certainty, but this must be balanced against computational considerations. When an element of $\bm{p}$ is close to zero in the posterior, this can cause problems for NUTS. See Section~\ref{sec:discuss} for a discussion of this issue. As a result, we use a value of $t=1/10$ which regularizes $\bm{p}$ away from zero.

For the $\alpha_k$s, we restrict the prior mass to be above one so that the untruncated Pareto distribution in the rightmost bin has a well defined mean. Note that $\alpha_k > 2$ is necessary to ensure a well defined variance if a user wants to include estimates of the second moment of the income distribution in the data model. Nevertheless, we assume that the $\alpha_k$s are iid truncated normal distributed as
\begin{align*}
  \alpha_k \stackrel{iid}{\sim}\mathrm{N}(2, 1^2)\mathbbm{1}(\alpha_k > 1).
\end{align*}
In practice we have found using PRLN that the tail bins tend to have estimated $\alpha_k$s between around one and three, with smaller values in bins further in the right tail. In general, there is not much information in the data to learn the Pareto parameters, so this prior provides some useful regularization to help with model estimation.

\section{EVALUATION OF L-PRLN}\label{sec:proof}
As mentioned above, the goal of the L-PRLN methodology as presented here is to provide estimates of income segregation indices and their uncertainty for ACS data, taking into account the survey errors and multiple data sources. Although the standard PRLN methodology cannot account for multiple and uncertain data sources, and only provides point estimates, it is useful to see how the L-PRLN approach compares to PRLN when considering only point estimates of income distributions.  Thus, we consider two specific cases. First, in Section~\ref{sec:sim}, we design a simulation study using a synthetic population generated over the Boone County, MO PUMA (Public-use Micro Area) and its census tracts. We repeatedly sample from this population and create synthetic tract-level ACS estimates, which we use to fit both PRLN and L-PRLN, and then evaluate them based on predictions of various features of the tract-level distributions. Then, in Section~\ref{sec:acs.test}, we use our modeling framework to estimate U.S. census tract-level income distributions using 2015 ACS 5-year period estimates associated with features of tract-level income distributions, and compare these to held out estimates to evaluate PRLN and L-PRLN.

Both of these exercises are designed to make a ``fair'' comparison between PRLN and L-PRLN. That is, recall that compared to PRLN, L-PRLN is able to use more of the available estimates, account for the uncertainty in those estimates, and provide uncertainty quantification. These are necessary properties to solve our income segregation problem. However, it is important to emphasize that here we compare the performance of L-PRLN to PRLN based only on point estimates, and restrict L-PRLN to use a much more limited subset of the estimates than it is capable of incorporating.

\subsection{Simulation study}\label{sec:sim}
We construct a synthetic population for our simulation study, and repeatedly sample from it using a stratified random sample based on the strata defined by the 2014 PUMS. We do not fully describe how the synthetic population is generated here; instead, see Appendix~\ref{app:synpop} of the Supplementary Material for a detailed description. Additionally, the \verb0R0 code \citep{R2017} used to generate the population is included in the Supplementary Material. Figure~\ref{fig:pops} in the Supplementary Materials contains maps of the true tract-level means, medians, and standard deviations of income for the synthetic population.

Similar to the real ACS, approximately 10\% of the population is sampled without replacement, and the sample size of each stratum is proportional to its sample size in the PUMS. Then the synthetic ACS estimates are created using the sample and associated weights in each tract, and the associated standard errors are created using successive difference replication \citep{judkins1990fay,fay1995aspects}, the method used in the ACS \citep{repwt2015,varrepwt}. We construct bin estimates, median estimates, and mean estimates in order to fit the models. We use the same 12 bin estimates that are available in the ACS, defined by the following breaks: $\$5,000$, $\$10,000$, $\$15,000$, $\$20,000$, $\$25,000$, $\$35,000$, $\$50,000$, $\$75,000$, $\$100,000$, $\$150,000$, and $\$200,000$. We also construct each fifth percentile estimate (5th, 10th, etc.) as well as the Gini index so that we can compare them to model-based estimates of the same quantities.

Then we fit L-PRLN using \verb0Rstan0 \citep{rstan2016} to do MCMC via NUTS with four chains, and after a warm-up of 4,000 iterations per chain for tuning and burn-in, a further 4,000 iterations per chain were kept as draws from the posterior distribution. Both the mean and the median of the posterior predictive distribution for each percentile were taken as model-based estimates. Additionally, we fit PRLN on the synthetic bin estimates. This yields four estimates of each percentile: the mean and median of the L-PRLN posterior predictive distribution, constructed as in Section~\ref{sec:post.pred}; the PRLN estimate; and the direct estimate. We computed the following four metrics for all four estimates: root mean square error (RMSE), mean absolute deviation (MAD), root mean square percentage error (RMSPE), and mean absolute percentage error (MAPE). All four metrics were computed over all iterations of the simulation study and all tracts of the synthetic population simultaneously.

\begin{table}[ht]
\centering
{\footnotesize
\begin{tabular}{llrrrrrrrrrr}
      \hline
      & Estimator & P5 & P10 & P15 & P20 & P25 & P30 & P35 & P40 & P45 & P50 \\
      \hline
      MAD & P. Mean & 8.58 & 15.38 & 17.86 & 15.24 & 14.80 & 11.08 & -2.00 & -10.45 & -11.25 & -7.21 \\
      & P. Median & 6.67 & 10.64 & 12.02 & 10.20 & 11.20 & 9.51 & 0.91 & -6.27 & -8.25 & -5.63 \\
      & PRLN & 1.55 & -1.41 & -2.31 & -4.80 & -2.94 & -1.80 & -3.03 & -4.81 & -4.00 & -1.15 \\ \hline
      MAPE & P. Mean & 3.90 & 12.99 & 15.37 & 12.14 & 13.17 & 10.51 & -0.43 & -7.98 & -9.96 & -6.44 \\
      & P. Median & 3.50 & 8.40 & 9.60 & 6.86 & 9.98 & 8.44 & 2.25 & -4.01 & -6.84 & -4.69 \\
      & PRLN & -0.83 & -1.35 & -2.22 & -5.62 & -2.42 & -2.39 & -2.73 & -4.29 & -3.92 & -1.02 \\ \hline
      RMSE & P. Mean & 7.61 & 17.94 & 23.53 & 20.07 & 13.43 & 7.09 & -4.98 & -11.71 & -12.43 & -9.78 \\
      & P. Median & 6.43 & 12.86 & 17.14 & 15.59 & 11.05 & 7.20 & -1.60 & -7.32 & -9.29 & -7.77 \\
      & PRLN & -0.60 & -2.98 & -2.36 & -3.94 & -3.75 & -2.95 & -4.41 & -4.76 & -3.88 & -2.17 \\ \hline
      RMSPE & P. Mean & 2.67 & 14.64 & 19.55 & 16.63 & 12.91 & 8.11 & -1.35 & -7.34 & -9.91 & -8.22 \\
      & P. Median & 2.73 & 9.82 & 13.43 & 11.81 & 10.24 & 7.64 & 1.43 & -3.27 & -6.74 & -6.05 \\
      & PRLN & -3.34 & -2.59 & -2.10 & -4.31 & -3.28 & -3.18 & -3.83 & -4.03 & -3.68 & -1.97 \\
      \hline
\end{tabular}
  }
\caption{Percentage difference in a variety of metrics between several estimates and the direct estimates for the first half of the income distribution. The estimates considered include the original Pareto-linear procedure (PRLN) the posterior predictive mean from L-PRLN (P. Mean), and the posterior predictive median from L-PRLN (P. Median). Negative numbers indicate that the method is doing better than the direct estimates.}
\label{tab:sim}
\end{table}

\begin{table}[ht]
  \centering
  {\footnotesize
    \begin{tabular}{llrrrrrrrrrr}
      \hline
      & Estimator & P55 & P60 & P65 & P70 & P75 & P80 & P85 & P90 & P95 & Gini \\
      \hline
      MAD & P. Mean & -3.95 & -0.08 & -0.60 & 0.91 & 2.56 & 7.38 & 3.52 & -1.61 & -1.71 & 14.58 \\
      & P. Median & -2.49 & 1.85 & 1.75 & -0.08 & 5.03 & 10.49 & 6.25 & 0.87 & 5.07 & 12.14 \\
      & PRLN & -1.79 & -3.17 & -6.32 & -4.27 & -2.77 & -0.00 & -2.60 & 0.45 & 38.89 & 9.46 \\ \hline
      MAPE & P. Mean & -4.26 & -0.75 & -0.84 & -0.12 & 1.17 & 5.93 & 2.84 & -1.68 & -1.85 & 15.19 \\
      & P. Median & -2.80 & 0.92 & 1.96 & -1.20 & 3.91 & 8.59 & 5.46 & 0.85 & 4.93 & 12.68 \\
      & PRLN & -1.71 & -2.85 & -5.62 & -3.59 & -3.43 & -0.95 & -2.62 & 0.10 & 37.59 & 9.71 \\ \hline
      RMSE & P. Mean & -5.62 & -2.61 & -2.26 & -2.66 & -3.00 & 1.48 & -2.48 & -6.52 & -8.79 & 10.55 \\
      & P. Median & -3.52 & -0.20 & -0.40 & -2.50 & -1.25 & 4.08 & 0.26 & -3.66 & -3.91 & 9.17 \\
      & PRLN & -2.12 & -3.63 & -5.70 & -5.04 & -4.68 & -2.34 & -4.08 & 5.25 & 57.91 & 8.96 \\ \hline
      RMSPE & P. Mean & -5.95 & -3.69 & -2.89 & -4.17 & -4.57 & -0.47 & -3.56 & -6.66 & -9.26 & 11.19 \\
      & P. Median & -4.05 & -1.53 & -0.66 & -3.90 & -2.53 & 1.70 & -1.11 & -3.82 & -4.36 & 9.74 \\
      & PRLN & -2.05 & -3.12 & -4.51 & -4.03 & -5.43 & -3.53 & -4.34 & 3.76 & 53.92 & 9.35 \\
      \hline
    \end{tabular}
  }
\caption{Percentage difference in a variety of metrics between several estimates and the direct estimates for the last half of the income distribution and the Gini coefficient. The estimates considered include the original Pareto-linear procedure (PRLN) the posterior predictive mean from L-PRLN (P. Mean), and the posterior predictive median from L-PRLN (P. Median). Negative numbers indicate that the method is doing better than the direct estimates.}
\label{tab:sim2}
\end{table}

Tables~\ref{tab:sim}~and~\ref{tab:sim2} display each of these metrics, expressed as a percentage of the same metric for the corresponding direct estimates. For example, PRLN had an RMSE for the 5th percentile 0.60\% lower than that of the direct estimate, while it had a MAD for the 5th percentile 1.55\% higher than that of the direct estimate. Note that the direct estimates are what our hypothetical data user would like the ACS to publish, but they are not available.

In the lower portion of the income distribution, the PRLN estimate does the best according to most metrics, while the L-PRLN posterior median outperforms the posterior mean. In the middle of the distribution this completely reverses: PRLN does the worst, and the posterior mean outperforms the posterior median. In the upper portion of the distribution but still under the 90th percentile, PRLN does the best again, but the posterior mean still outperforms the posterior median. In the 90th percentile, the posterior mean performs the best, while PRLN performs the worst. In the 95th percentile the same pattern holds, but PRLN performs disastrously bad. This is because if PRLN cannot guarantee an estimate for an $\alpha$ that is greater than one in the top bin, it assumes the bin is a point mass on the bin minimum. See Section~\ref{sec:PRLN} for details. This can drastically hurt PRLN's predictions in the upper tail, which we see here. L-PRLN does not have this problem since each $\alpha$ is constrained to be greater than one and is regularized away from one by the prior.

So in general, the best performing point-estimate depends on which region of the income distribution the data-user cares about. For the middle of the distribution or the far right tail, L-PRLN is superior, but everywhere else PRLN is superior. PRLN performs the best for the Gini coefficient, with the posterior median outperforming the posterior mean. For other measures of inequality and other functionals of the income distributions, which estimate performs best will depend on how much they load on different regions of the income distribution. Note that this comparison deliberately limited L-PRLN by preventing it from using all of the available estimates -- estimates that PRLN cannot use.

It is also important to emphasize that L-PRLN provides uncertainty estimates, which are unavailable in PRLN. As an illustration, Table~\ref{tab:cover} presents the coverage rates of 95\% credible intervals for every fifth percentile, as well as the Gini coefficient. Two coverage rates were computed, one with the true population as reference values and one with the PRLN estimates as reference values. The first comparison shows that L-PRLN's intervals slightly undercover the truth; i.e., the 95\% credible intervals cover about 80-90\% of the time, but with better coverage in the lower portion of the income distribution. Note that L-PRLN has better coverage precisely where its point estimates do the worst. The second comparison shows that the PRLN measures in the lower part of the distribution are largely contained in the L-PRLN's 95\% credible intervals. More precisely, L-PRLN's estimate and PRLN's estimate for a given percentile were statistically indistinguishable at least 60\% of the time. This is an underestimate since it does not account for uncertainty in the PRLN estimates, but the statistical properties of PRLN are unknown.

\begin{table}[ht]
  \centering
  \begin{tabular}{lrr|lrr}
    \hline
    Estimand & Population & PRLN & Estimand & Population & PRLN \\
    \hline
    P5  & 0.92 & 0.86 &    P55 & 0.79 & 0.60 \\
    P10 & 0.90 & 0.81 &    P60 & 0.79 & 0.62 \\
    P15 & 0.89 & 0.78 &    P65 & 0.82 & 0.69 \\
    P20 & 0.90 & 0.75 &    P70 & 0.80 & 0.71 \\
    P25 & 0.90 & 0.74 &    P75 & 0.80 & 0.72 \\
    P30 & 0.89 & 0.73 &    P80 & 0.81 & 0.73 \\
    P35 & 0.88 & 0.71 &    P85 & 0.85 & 0.76 \\
    P40 & 0.87 & 0.69 &    P90 & 0.85 & 0.79 \\
    P45 & 0.85 & 0.65 &    P95 & 0.88 & 0.75 \\
    P50 & 0.82 & 0.63 &    Gini & 0.95 & 0.84 \\
    \hline
\end{tabular}
\caption{Coverage rates of 95\% credible intervals from the tract level model for each quantity of interest, averaged over tracts. Coverage rates are computed taking the true population value as the reference value (Population), and taking the PRLN estimate as the reference value (PRLN).}
\label{tab:cover}
\end{table}

\subsection{Application to the American Community Survey}\label{sec:acs.test}
We fit PRLN and L-PRLN to 2015 ACS 5-year period estimates of features of tract-level income distributions for all tracts in five separate PUMAs: PUMA 821 in Colorado (a wealthy rural PUMA south of Denver), PUMA 3502 in Illinois (a wealthy PUMA in the northern portion of Chicago), PUMA 600 in Missouri (Boone County, MO, a college town and rural outlying areas), PUMA 600 in Montana (a sparsely populated rural PUMA), and 3706 in New York (a poor urban PUMA in New York City). Figure~\ref{fig:allpumas.median} in the Supplementary Materials contains maps of each PUMA and each of their Census tracts, shaded according to the 2015 ACS 5-year period estimate of median household income.

We fit the models using each of the bin estimates described in Section~\ref{sec:acs.model}, as well as the mean and median estimates. We held out estimates of the 20th, 40th, 60th, 80th, and 95th percentile, as well as the Gini coefficient to validate the models. To fit each model we used \verb0Rstan0 \citep{rstan2016} to do MCMC via NUTS with four chains, a warm-up of 4,000 iterations per chain for tuning and burn-in, and a further 4,000 iterations per chain were kept as draws from each model's posterior distribution.

For L-PRLN, we construct the posterior predictive mean and median for each estimand, as in Section~\ref{sec:post.pred}. We compare each of these estimates as well as estimates from PRLN to each of the held out estimates using the same four metrics as in Section~\ref{sec:sim}: RMSE, RMSPE, MAD, and MAPE, all computed across tracts. Tables~\ref{tab:co.metric}--\ref{tab:ny.metric} of the Supplementary Materials contain these metrics for each of the five PUMAs we considered. Note that for some tracts, some of the held out estimates were missing -- particularly the 95th percentile, and mainly in the IL PUMA.

For most estimands in most tracts, and according to most metrics, L-PRLN does about the same or slightly worse than PRLN. The main exceptions are in either tail of the distribution, where for some tracts the difference between PRLN and L-PRLN is more magnified. L-PRLN especially has trouble relative to PRLN in the lower tail. On the other hand, L-PRLN often performs better than PRLN for the Gini coefficient, and in particular in the IL PUMA it performs much better for the 95th percentile and consequently for the Gini coefficient. This is due to the phenomenon discussed in Section~\ref{sec:sim}, where PRLN sometimes significantly incorrectly estimates the distribution in the upper bin. Additionally, in the CO PUMA, the L-PRLN outperforms PRLN in the middle of the distribution. 

\section{INCOME SEGREGATION INDICES}\label{sec:index}
Now we turn to our motivating problem: estimating income segregation indices using ACS data. Households are segregated by income to the extent that households with similar incomes choose to live near each other. To measure this, \citet{reardon2011measures,reardon2011income} construct the rank-order information theory index. The basic idea of the index is to construct an entropy measure of the income distribution for an entire metro area, and then construct the same measure for the income distributions for each census tract in the metro area. Then the index is a weighted sum of the relative differences in this entropy measure between each tract and the metro area.

Formally, let $F_i(y)$ denote the CDF of income for census tract $i$, where $i=1,2,\dots,I$ indexes all census tracts in a given metro area. Then we assume that the income distribution for the metro area, denoted by $F(y) = \sum_{i=1}^Iw_iF_i(y)$, is a population weighted mixture of the tract-level income distributions, where $w_i$ is the proportion of the metro area's population in tract $i$. Next, define $E(G||F)$ as the integrated binary entropy from the CDF F to the CDF G, i.e.
\begin{align}
E(G||F) = \int_{-\infty}^{\infty}e[F(y)]dG(y) \label{eq:int.entropy}
\end{align}
where $e(p) = -p\log(p) -(1-p)\log(1-p)$ is binary entropy. Then the rank-order information theory index, denoted by $H_R$, can be defined as
\begin{align}
H_R = \sum_{i=1}^Iw_i\frac{E(F||F) - E(F||F_i)}{E(F||F)} \label{eq:info}.
\end{align}
Since $H_R$ is based on entropy, it is better understood as a measure of the differences in \emph{diversity} of the income distributions between the tract-level and metro-level \citep{roberto2015divergence}. Indeed, the following example illustrates the point. Suppose that households in the metro area only have one of two incomes: $y=30,000$ and $y=100,000$. In the entire metro area $P(y=30,000) = 2/3$, while in tract $i$, $P_i(y=30,000) = 1/3$. Then for tract $i$ we have
\begin{align*}
  E(F_i||F) &= -e[F_i(30,000)] P(y=30,000) - e[F_i(100,000)] P(y=100,000) \\
            &= -e[1/3] \frac{2}{3} - e[1] \frac{1}{3} = -e[2/3] \frac{2}{3} = E(F||F)
\end{align*}
since $e(p) = e(1-p)$. So tract $i$ contributes nothing to the metro area's segregation index even though it has a much higher concentration of rich households than the entire metro area.

To remedy this, \citet{roberto2015divergence} proposes the KL divergence index. Let $f$ denote the PDF associated with $F$ above, and similarly for $f_i$ and $F_i$. Then the KL divergence index can be defined as
  \begin{align}
    D &= \sum_{i=1}^Iw_i D(f_i||f), &&& D(g||f)&= \int_{-\infty}^{\infty}\log \frac{g(y)}{f(y)}g(y)dy \label{eq:div}
  \end{align}
  where $D(g||f)$ is the KL divergence from the PDF $f$ to the PDF $g$. In other words, the divergence index is the population weighted sum of the divergences from the metro-level income distribution to each of the tract-level distributions.

\subsection{Correlates of income segregation}\label{sec:correlates}
\citet{reardon2011income} investigate the correlates of income segregation as measured by $H_R$. In particular, they are interested in whether income inequality, as measured by the Gini index, is correlated with income segregation. They consider the largest 100 metro areas in the U.S. by population, and fit a variety of regression models controlling for various covariates. We focus on a portion of their Table~4, which reports the results of severale regression models, of which we focus on three: one for all families, one for black families only, and one for white families only. In these models they control for the year of the census, various metro-year and race-metro-year covariates, and include metro fixed effects. They find a stable positive relationship between the Gini coefficient and $H_R$. Further, the strength of this relationship is about the same for white families alone as it is for black families alone.  Our aim is to attempt to rerun these regressions using recent ACS data, then run them again replacing $H_R$ with the divergence index.

Since we use ACS data, our controls and data differ in several ways in general. First, we use the top 100 metro areas by population according to the 2018 ACS 5-year period estimates of population. This list may not be identical to the list used by \citet{reardon2011income}. Second, we use ACS estimates for households instead of families because more of the required variables are available, though \citet{reardon2011income} note that they would have preferred to do a household level analysis, but it was not possible due to data limitations. Finally, we only use a single year of ACS 5-year period estimates. The ACS is not old enough to have more than two years of non-overlapping 5-year period estimates. We use the 2018 5-year period estimates. 2013 5-year period estimates are also available, though the definitions of several covariates differ across vintages. To avoid this complication, we only use a single year of estimates. This leaves us with no within-metro variation in any of the three regression models, so we omit the metro area fixed effects. Otherwise, we attempt to faithfully include every covariate in the regression of \citet{reardon2011income} in our own regressions. Appendix~\ref{app:data} of the Supplementary Materials describes how each covariate was sourced from the 2018 ACS 5-year period estimates, including how standard errors were constructed if necessary. For black households and white households, two covariates are not available: percentage of households with a female householder, and the Gini index. We omit the female householder covariate in the black households and white households regressions for this reason. However, we take advantage of L-PRLN to construct the metro-level Gini index for black households and white households only along with its standard error using the available metro-race-level income estimates. See Appendix~\ref{app:gini} for a description of how this was performed. 

To construct both $H_R$ and $D$ for a given metro area, we first fit L-PRLN to the household ACS 5-year period estimates of variations features of the household income distribution for each tract in that metro area. For the household income distributions, we use bin estimates with the same boundaries as in Section~\ref{sec:sim}, mean and median estimates, estimates of the 20th, 40th, 60th, 80th, and 95th percentile, estimates of the income shares of the quintiles of the income distribution as well as the top 5\% of the income distribution, and an estimate of the Gini coefficient \citep{acs2018quintiles,acs2018shares,acs2018gini,acs2018income,acs2018financial}. For many tracts, some of these estimates or their standard errors are not available. For those tracts we proceed with whichever estimates with standard errors are available. In all cases if the mean estimate or its standard error was not available for the tract, or if the ACS estimate of the population of households was less than 100, the tract was omitted from the analysis. The same procedure was applied to estimating the tract-level income distributions of black households alone, and of white households alone, again as long as there were at least 100 households of the given race in the tract according to the ACS. The only available tract-race-level household income distribution ACS estimates are the bin, mean, and median estimates. The same priors as in Section~\ref{subsec:priors} were used, except in the black households models, country-level bin estimates for only black households were used to center the prior on the bin probabilities, and similarly for the white households models.

We cannot use the approach in Section~\ref{sec:post.pred} to estimate $H_R$ and $D$ using the L-PRLN income distribution estimates because both $H_R$ and $D$ will yield nonsensical results if each tract's income distribution does not have the same support. So instead we treat both indices as a function of the underlying tract-level parameters. Then we approximate the integrals in \eqref{eq:int.entropy} and \eqref{eq:div} for each draw from the posterior distribution using importance sampling techniques -- see Appendix~\ref{app:compute.indices} for details. The result of this process is that for a given metro area, we obtain a joint posterior sample of the index and the standard error associated with approximating the integrals. Our approach is the same for computing $H_R$ and $D$ by race in a given metro area.

\subsection{Error-in-variables regression}\label{sec:eiv}
To fit the regressions, we must contend with two complications that were not present in \citet{reardon2011income}. First, the response and each covariate of each regression is measured with error, though in each case the standard error is known. Second, instead of observing the response and its standard error, we observe a sample from the joint posterior distribution of the response and its standard error. 

The solution to the first issue is an EIV regression; e.g., see \citet{carroll2006measurement}, \citet{arima2015bayesian}, and the references therein. But to use that, we first need to solve the second problem using the variance decomposition formula. Let $\bm{\theta}$ denote all unknown parameters of the tract-level L-PRLN models for a given metro area, let $d^*(\bm{\theta})$ denote a segregation index as a function of those parameters, let $d$ denote our estimate of that index, and let $h(\bm{\theta})$ denote the corresponding standard error. Then conditional on the model parameters we have $d|\bm{\theta} \sim \mathrm{N}(d^*(\bm{\theta}), h^2(\bm{\theta}))$.
Then we can write $\E[d] = \E[d^*(\bm{\theta})]$ and $\var[d] = \E[h^2(\bm{\theta})] + \var[d^*(\bm{\theta})]$.
Given a sample $\{(d_m, h_m^2): m=1,2,\dots,M\}$ from the joint posterior of $(d, h^2(\bm{\theta}))$, we can approximate these quantities by
\begin{align*}
  \E[d] &\approx \overline{d} = \frac{1}{M}\sum_{m=1}^Md_m\\
  \var[d] &\approx \overline{h}^2 = \frac{1}{M}\sum_{m=1}^Mh^2_m + \frac{1}{M-1}\sum_{m=1}^M(d_m - \overline{d})^2.
\end{align*}
Then for simplicity we assume a simple measurement error model using these quantities:
\begin{align*}
\overline{d} \sim \mathrm{N}(d^*, \overline{h}^2),
\end{align*}
where $d^*$ is the true underlying index. This approach works for both $H_R$ and $D$, and allows us to completely reduce the regression problem to EIV regression.

The EIV regression model we employ can be written as follows. Let $i=1,2,\dots,I$ index metro areas, let $\overline{d}_i$ denote either segregation index for that metro area, and let $\overline{h}_i$ denote its associated standard error. Let $\bm{x}_i$ denote a vector of covariates for the metro area, with $\bm{S}_i$ the associated (diagonal) error covariance matrix. Further, let $d_i^*$ and $\bm{x}_i^*$ denote the latent true values of the index and covariates, respectively. Then the model is given by \eqref{eq:eivreg}
\begin{align}
  \overline{d}_i | \bm{x}_i, d_i^*, \bm{x}_i^* &\sim \mathrm{N}(d_i^*, \overline{h}_i^2) \nonumber\\
  \bm{x}_i | d_i^*, \bm{x}_i^* &\sim \mathrm{N}(\bm{x}_i^*, \bm{S}_i) \nonumber\\
  d_i^* | \bm{x}^*_i & \sim \mathrm{N}(\alpha + (\bm{x}_i^*)'\bm{\beta}, \tau^2)\nonumber\\
  x_{ij}^* & \stackrel{ind}{\sim} \mathrm{N}(\mu_j, \sigma_j^2), \label{eq:eivreg}
\end{align}
for $i=1,2,\dots,I$, where $j=1,2\dots,J$ indexes covariates. To complete the model we need priors on $\alpha$, $\bm{\beta}$, $\tau^2$, the $\mu_j$s, and the $\sigma_j^2$s.

We specify priors on the standardized regression coefficients for ease of interpretation and elicitation, and on the corresponding standardized versions of all other parameters, i.e. on $\widetilde{\beta}_j = \beta_j s_{x_j} / s_{\overline{d}}$, $\widetilde{\mu}_j = (\mu_j - \overline{x}_j)/s_{x_j}$, and $\widetilde{\sigma}_j = \sigma_j / s_{x_j}$ for $j=1,2\dots,J$, and on $\widetilde{\alpha} = (\alpha - \overline{\overline{d}} + \overline{\bm{x}}'\bm{\beta}) / s_{\overline{d}}$ and $\widetilde{\tau} = \tau / s_{\overline{d}}$. Using this parameterization, we employ the independent priors listed in \eqref{eq:eivprior} 
\begin{align}
  \widetilde{\alpha} &\sim \mathrm{N}(0, 100^2) &&& \nonumber\\
  \widetilde{\beta}_j &\stackrel{iid}{\sim} \mathrm{N}(0, 3^2) &&& \mbox{ for } j=1,2,\dots,J \nonumber\\
  \widetilde{\tau} & \sim \mathrm{N}^+(0, 0.8^2) &&&\nonumber\\
  \widetilde{\mu}_j & \stackrel{iid}{\sim} \mathrm{N}(0, 3^2) &&& \mbox{ for } j=1,2,\dots,J\nonumber\\
  \widetilde{\sigma}_j & \stackrel{iid}{\sim} \mathrm{N}^+(0, 2^2) &&& \mbox{ for } j=1,2,\dots,J. \label{eq:eivprior}
\end{align}
The priors on the $\widetilde{\beta}_j$s imply that for any covariate, we are 68\% sure that a one standard deviation change in the covariate will result in no more than a three standard deviation change in the response. The prior on $\widetilde{\alpha}$ implies that we are 68\% certain that the intercept will be within 100 sample standard deviations of the response from the sample mean of the response. The half-normal prior on $\widetilde{\tau}$ implies that we are 68\% certain that the error variance will be no more than $0.8$ times the the total sample variance of the response. All of these priors are loose relative to typical expectation of regressions in the social sciences, but still provide a small amount of regularization.

The priors on the covariate means and standard deviations are similarly loose, and can be thought of as empirical Bayes priors. The priors on the $\mu_j$s are loosely centered on the sample means of the $x_j$s, and the priors on the $\sigma_j$s allow for a wide range of variation around the sample standard deviations of the $x_j$s.

\subsection{Results}\label{sec:results}
Figures~\ref{fig:indexplot}, \ref{fig:blackindexplot}, and \ref{fig:whiteindexplot} in Appendix~\ref{app:eiv.results} demonstrate that the divergence and information theory indices substantially disagree about the relative ranking of metro areas in terms of income segregation. This demonstrates that \citet{roberto2015divergence}'s criticism of the information theory index is not merely a theoretical curiosity, but instead that there is significant mismeasurement of income segregation. As a result, we should expect the EIV regression results to differ as well.

Table~\ref{tab:eiv.results.raw} contains posterior summaries of the Gini index EIV regression coefficient for each model fit --- see Appendix~\ref{app:eiv.results} for detailed tables including every covariate. In similar information index regressions, \citet[Table 4]{reardon2011income} found that the regression coefficient on the Gini index to be $0.56$ for all families, $0.47$ for black families, and $0.47$ for white families. Our regressions use more recent household-level data, do not have exactly the same covariates, and do not have metro fixed effects. Despite this, our results for all households are broadly consistent with the results of \citet{reardon2011income} for all families. Our results for black and white households are somewhat different. In both cases the 95\% credible intervals are much wider, and contain zero. This is largely due to higher standard errors for the black households and white households ACS estimates compared to the standard errors of the all households ACS estimates. The covariates simply contain less useful information for the black households and white households regressions.

However, the Gini index regression coefficients also appear to be much closer to zero for black households and white households. The upper end of the 95\% credible intervals do not contain \citet{reardon2011income}'s estimates, and for black households the posterior mean and median are both negative. This may be due to differences in model specification since our black households and all households regressions are missing the female head of household covariate since it is not available in the ACS. Additionally, our regressions do not include metro area fixed effects since we have only one year of data, though this difference is present in the all households regressions as well. That said, we take these regressions as a baseline to compare with divergence index regressions using the same model specification.

\begin{table}[ht]
  \centering
  \begin{tabular}{rrrrrrrr}
    \hline
    Households & Mean & SD & 2.5\% & 25\% & 50\% & 75\% & 97.5\% \\ 
    \hline
    \multicolumn{8}{l}{Information theory index} \\
    All   &  0.427 & 0.058 &  0.313 &  0.389 &  0.427 & 0.467 & 0.540 \\ 
    Black & -0.079 & 0.125 & -0.328 & -0.161 & -0.078 & 0.004 & 0.164 \\ 
    White &  0.145 & 0.111 & -0.072 &  0.071 &  0.146 & 0.220 & 0.366 \\ 
    \multicolumn{8}{l}{Divergence index} \\  
    All   & 0.763 & 0.157 &  0.458 & 0.657 & 0.762 & 0.868 & 1.074 \\ 
    Black & 1.403 & 0.742 & -0.036 & 0.899 & 1.402 & 1.898 & 2.872 \\ 
    White & 0.683 & 0.240 &  0.207 & 0.523 & 0.683 & 0.843 & 1.155 \\ 
    \hline
  \end{tabular}
  \caption{Posterior summaries of raw EIV regression coefficients for the Gini index.}
  \label{tab:eiv.results.raw}
\end{table}

The results for the divergence index are significantly different. The coefficient on the Gini index is larger in all cases, though for making these comparisons the standardized coefficients displayed in Table~\ref{tab:eiv.results.std} is more meaningful. In that case, the Gini index regression coefficient is similarly sized in the all households regressions, though there is more uncertainty in the divergence index regression. The Gini index coefficients in the black households and white households regressions are once again smaller than in the all households regressions, but this difference is much less extreme for the divergence index than for the information theory index. In fact, the 95\% credible interval for white households is strictly above zero, and for black households only just contains zero inside the lower bound. The upshot is that the Gini index appears to be positively associated with the divergence index for black households only and white households only, while the same is not true for the information theory index. That said, there is enough uncertainty that we cannot rule out that there is no meaningful difference between the divergence index coefficients and the information theory index coefficients.

\begin{table}[ht]
  \centering
  \begin{tabular}{rrrrrrrr}
    \hline
    Households & Mean & SD & 2.5\% & 25\% & 50\% & 75\% & 97.5\% \\ 
    \hline
    \multicolumn{8}{l}{Information theory index} \\
    All   &  0.566 & 0.076 &  0.414 &  0.515 &  0.565 & 0.618 & 0.714 \\ 
    Black & -0.100 & 0.157 & -0.412 & -0.203 & -0.098 & 0.005 & 0.207 \\ 
    White &  0.138 & 0.106 & -0.069 &  0.067 &  0.139 & 0.209 & 0.348 \\ 
    \multicolumn{8}{l}{Divergence index} \\
    All   & 0.580 & 0.120 &  0.348 & 0.500 & 0.580 & 0.661 & 0.817 \\ 
    Black & 0.379 & 0.200 & -0.010 & 0.243 & 0.378 & 0.512 & 0.775 \\ 
    White & 0.399 & 0.140 &  0.121 & 0.306 & 0.399 & 0.493 & 0.675 \\ 
    \hline
  \end{tabular}
  \caption{Posterior summaries of standardized EIV regression coefficients for the Gini index.}
  \label{tab:eiv.results.std}
\end{table}

\section{DISCUSSION}\label{sec:discuss}
L-PRLN serves its purposes well. It interpolates the income distribution nearly as well as the original PRLN when forced to use a restricted subset of the available estimates. However, it has several added benefits. First, our L-PRLN is able to take advantage of a wider variety of tract-level estimates than PRLN, including quantile and moment estimates. PRLN is fundamentally limited to using only bin estimates. Second, unlike PRLN, L-PRLN takes into account the standard errors of the tract-level estimates. Finally, while PRLN can only provide point estimates, L-PRLN provides uncertainty quantification through the posterior distribution.

While we employ L-PRLN to construct income segregation indices, it can be used to construct any other feature of income distributions of interest. For example, sociologists and economists are interested in a variety of measures of income ineqality and income segregation, and use a variety of methods to estimate them not limited to PRLN \citep{kennedy1996income,jargowsky1996take,mayer2001growth,hardman2004neighbors,watson2009inequality}. These approaches tend to suffer from the same limitations as PRLN, and  L-PRLN can be applied to estimating them as well.

L-PRLN can also be generalized and applied to other types of variables. For example, it could be used to interpolate the age distribution, for which there are often a selection of bin estimates available. To do this only requires appropriate choices for the $f_k$s in \eqref{eq:density.model}. Each $f_k$ could be a truncated normal density, though in practice the age distribution should be investigated to determine an appropriate choice. Many choices will require estimation of more parameters per bin than in the PRLN density. In order to handle this, it may be necessary to reduce the number of knots so that there are more bin estimates than knots. The framework can also be applied to data from sources other than the Census Bureau as well. The key is that there are a wide variety of available estimates of different distributional features at the area-level. These will typically be bin estimates, but many other estimate types could be used.

Based on the simulation study in Section~\ref{sec:sim} and out-of-sample performance on held out estimates in Section~\ref{sec:acs.test}, neither PRLN nor L-PRLN performed uniformly superior than the other when L-PRLN was restricted to a subset of available estimates. L-PRLN performed the best in the middle and far right tail of the distribution, with PRLN typically performing better elsewhere. This is likely due to how informative the Dirichlet prior is on the knot probabilities. As noted in Section~\ref{subsec:priors}, a more informative prior was necessary in this case to help facilitate NUTS.  In particular, note that for some census tracts, the bin estimate for one or more income categories is zero. Without an informative prior, these probabilities will be estimated to be close to zero and NUTS will go into the extreme tails of the transformed space, causing numerical and sampling problems. The informative prior regularizes those estimates away from zero and prevents the computational problem.  This leads to a loss of predictive accuracy, although this is reflected in the uncertainty estimates that are provided by L-PRLN.  Further, note the knots in L-PRLN are set equal to the boundaries defining the bins for the bin estimates.  This is done for computational convenience but is not necessary. Indeed, knot selection is a potential avenue for improving L-PRLN. Naively, it seems as though spacing the knots roughly equally in the quantile domain would alleviate the problem with probabilities being estimated close to zero, and improve the quality of the model. In model fits not reported here, we found that this degrades model performance despite the looser priors, suggesting that there are other factors important for knot selection. The number and spread of available tract-level estimates should fundamentally constrain the optimal number and placement of the knots in some way, but precisely how is an area of future research.

We turn now to the empirical application constructing income segregation indices and estimating the association between them and the Gini index. The analysis in \citet{reardon2011income} cannot be performed for more recent years due to the elimination of the decenial census long form. Instead, ACS estimates are available, but their standard error must be taken into account. Because of its deficiencies, PRLN is not suitable for constructing income segregation indices using ACS data. Unlike PRLN, L-PRLN allows us to use all available estimates, account for uncertainty in those estimates, and propagate that and other sources of uncertainty into the estimated indices. Additionally, L-PRLN played a secondary role by allowing us to construct metro-level Gini indices for black households only and white households only using ACS estimates, while again propagating uncertainty into the indices so that it could be accounted for in the EIV segregation index regressions. The segregation indices themselves disagreed substantially about the relative ranking of metro areas in terms of income segregation, illustrating that \citet{roberto2015divergence}'s criticisms of the information theory index are well-founded.

The results of the regressions were also instructive. Our regression results were meaningfully different when using the divergence index, though only for black households only and white households only. Using more recent ACS data and the information theory index in a somewhat different model specification, we were not able to reproduce \citet{reardon2011income}. Namely, that the Gini index and the information theory index were positively associated for both black households only and white households only, though there is a lot of uncertainty in our estimates. However, using the divergence index we do see a positive association with the Gini index for both black households only and white households only. The upshot is that we confirm one of the central conclusions of \citet{reardon2011income} -- that increased income inequality predicts increased income segregation even within racial groups -- but only with a proper measure of income segregation and not with their original index.

\begin{center}
{\large\bf SUPPLEMENTARY MATERIAL}
\end{center}

\begin{description}
\item[Online Appendix:] Includes several appendices adding relevant detail to the paper.
\begin{description}
\item[Appendix~\ref{app:explore}: Exploratory tables and figures.] Includes various tables and figures referenced throughout the paper that are useful, but not necessary, for understanding the data and results in this paper.
\item[Appendix~\ref{app:functionals}: Latent PRLN density functionals.] Includes formulas for all of the relevant functionals of the latent PRLN density referenced in the paper, including mean, variance, CDF, quantile function, income shares, and Gini index.
\item[Appendix~\ref{app:synpop}: Generating the synthetic population.] Includes details about how the synthetic population was generated in the simulation study in Section~\ref{sec:sim}.
  \item[Appendix~\ref{app:acstab}: Evaluating Point Estimates.] Includes tables evaluating L-PRLN and PRLN point estimates on a variety of metrics from comparison using ACS data in Section~\ref{sec:acs.test}.
\item[Appendix~\ref{app:data}: Segregation index EIV data.] Includes details on how the data for the segregation index EIV regressions were sourced from the ACS for Section~\ref{sec:index}.
\item[Appendix~\ref{app:gini}: Household level Gini index estimation by race.] Includes details on metro-level and metro-race-level household income Gini indices were estimated for use in Section~\ref{sec:index}.
\item[Appendix~\ref{app:compute.indices}: Computing segregation indices.] Includes details on how both the information theory index and the divergence index were computed as a function of model parameters in the posterior distribution for use in Section~\ref{sec:index}.
\item[Appendix~\ref{app:eiv.results}: Segregation index results.] Includes detailed tables of regression coefficients for each of the income segregation index EIV regressions we performed in Section~\ref{sec:index}.  
\end{description}
\item[Github Repository:] \url{https://www.github.com/simpsonm/latentprln} Includes all code for the all models discussed and used in the paper, and for reproducing our results, including \verb0R0 code for the Pareto-linear procedure, and code for downloading and cleaning ACS tables.
\end{description}

\clearpage
\newpage

\Appendix

\renewcommand{\thetable}{\Alph{section}.\arabic{table}}
\section{EXPLORATORY TABLES AND FIGURES}\label{app:explore}
\setcounter{table}{0}
This appendix contains several tables and figures that are useful for understanding the data that were referenced in the main text.

\begin{table}[h]
  \centering
  {\footnotesize
    \begin{tabular}{crrrrrrrrrr}
                & \multicolumn{10}{c}{Bins}\\\hline
          &       & $\geq$10 & $\geq$15 & $\geq$25 & $\geq$35 & $\geq$50 & $\geq$750 & $\geq$100 & $\geq$150 & $\geq$200\\
    Tract & $<$10 & $<$15 & $<$25 & $<$35 & $<$50 & $<$75 &$<$100 &$<$150 &$<$200 & \\
    \hline
 2      & 9.8  & 9.3  & 25.8 & 13.7 & 20.4 & 14.3 & 4.0  & 2.8  & 0.0  & 0.0\\
 3      & 31.9 & 16.0 & 21.1 & 12.4 & 3.3  & 6.8  & 4.1  & 1.9  & 1.1  & 1.4 \\
 5      & 46.6 & 8.3  & 19.5 & 6.4  & 10.3 & 3.8  & 1.7  & 0.9  & 2.5  & 0.0 \\
 6      & 7.2  & 3.2  & 4.4  & 3.6  & 16.1 & 17.3 & 14.2 & 23.0 & 5.8  & 5.4 \\
 7      & 10.5 & 10.8 & 15.3 & 15.7 & 16.6 & 18.9 & 9.1  & 2.7  & 0.4  & 0.0 \\
 9      & 17.6 & 10.3 & 21.5 & 14.6 & 18.4 & 10.4 & 4.9  & 2.2  & 0.0  & 0.0 \\
  \end{tabular}
}
\label{tab:acs}
\caption{Bin estimates for selected tracts in PUMA 600 (Boone County) in MO. All estimates are 2015 ACS 5-year period estimates, and come from ACS Table S1901. Each bin estimate is the percentage of households in that tract with an income within a set of bounds, including the lower bound but excluding the upper bound. Both bounds are denominated in \$1,000. The ACS tables also include an associated margin of error for each estimate (not displayed here).}
\end{table}

\begin{figure}
\centering
\includegraphics[scale = 0.9]{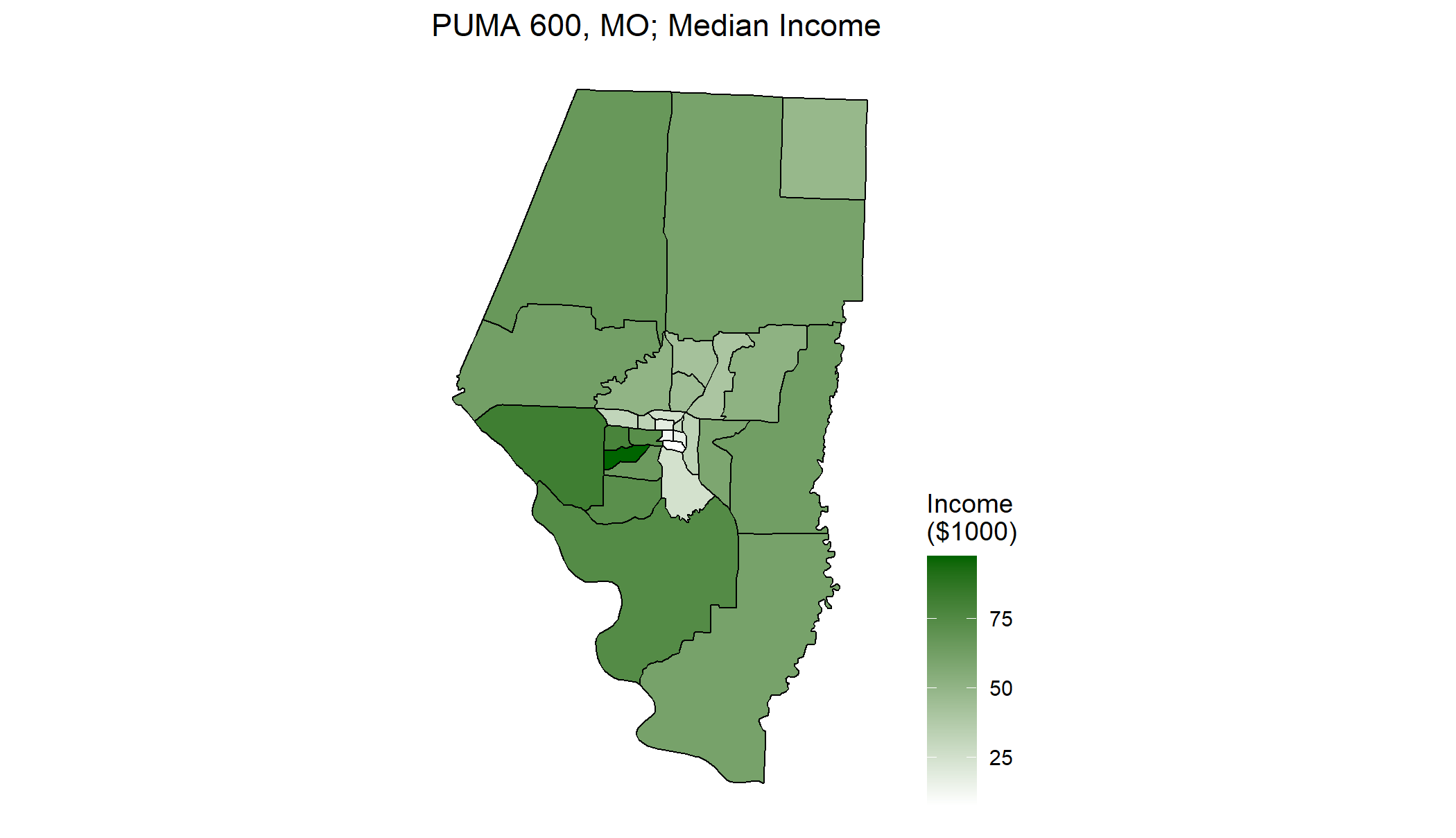}
\caption{An example PUMA with nested tracts: PUMA 600 (Boone County) in MO. Tracts are shaded according to 2015 ACS 5-year estimates of median household income.}
\label{fig:boonepuma}
\end{figure}

\begin{figure}
\centering
\includegraphics[width = 0.3\textwidth]{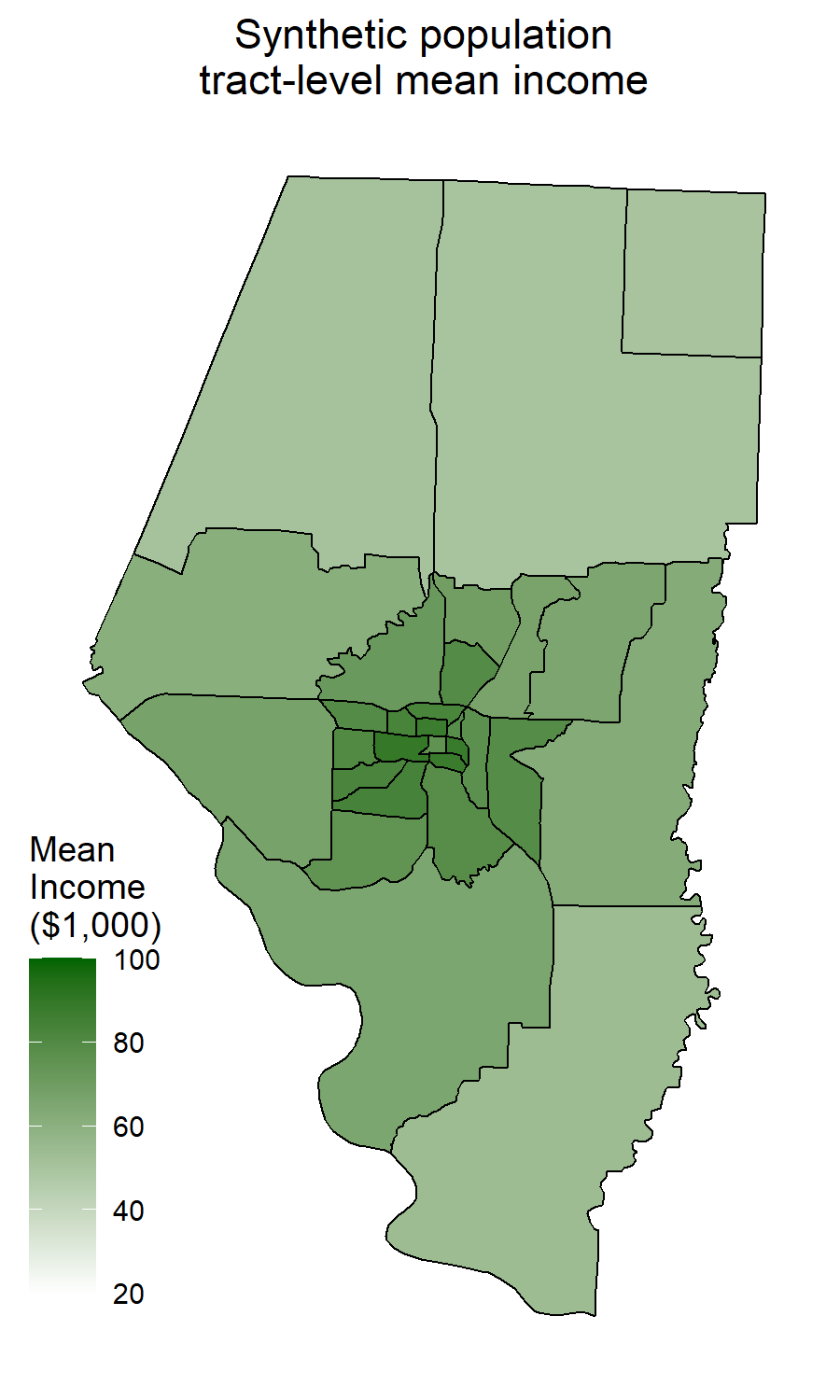}
\includegraphics[width = 0.3\textwidth]{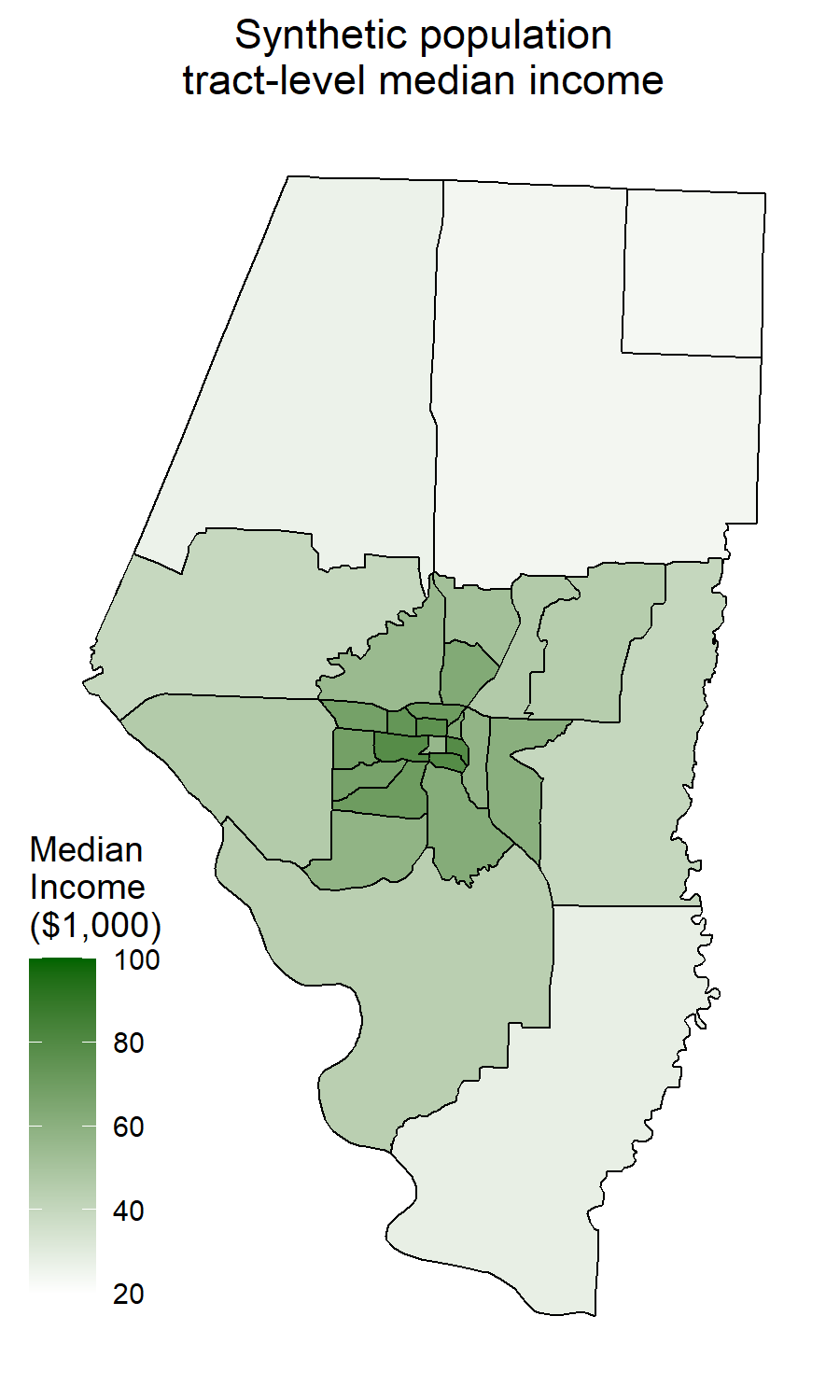}
\includegraphics[width = 0.3\textwidth]{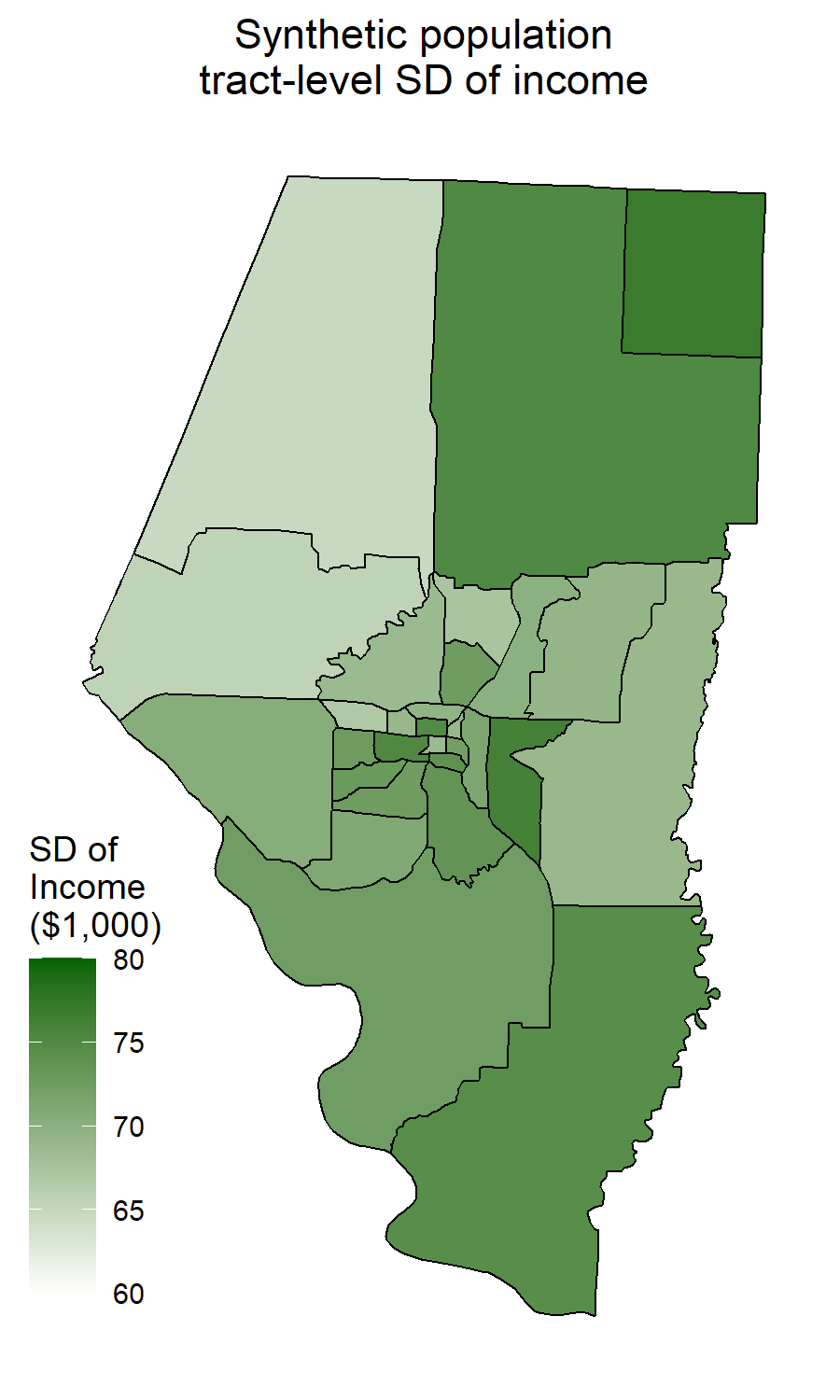}
\caption{True tract-level means, medians, and standard deviations of income for the synthetic population. The first two exhibit a noticeable inside-out spatial pattern, while the third is a bit different but still appears to have spatial dependence.}
\label{fig:pops}
\end{figure}

\begin{figure}
  \centering
  \includegraphics[width = 0.45\textwidth]{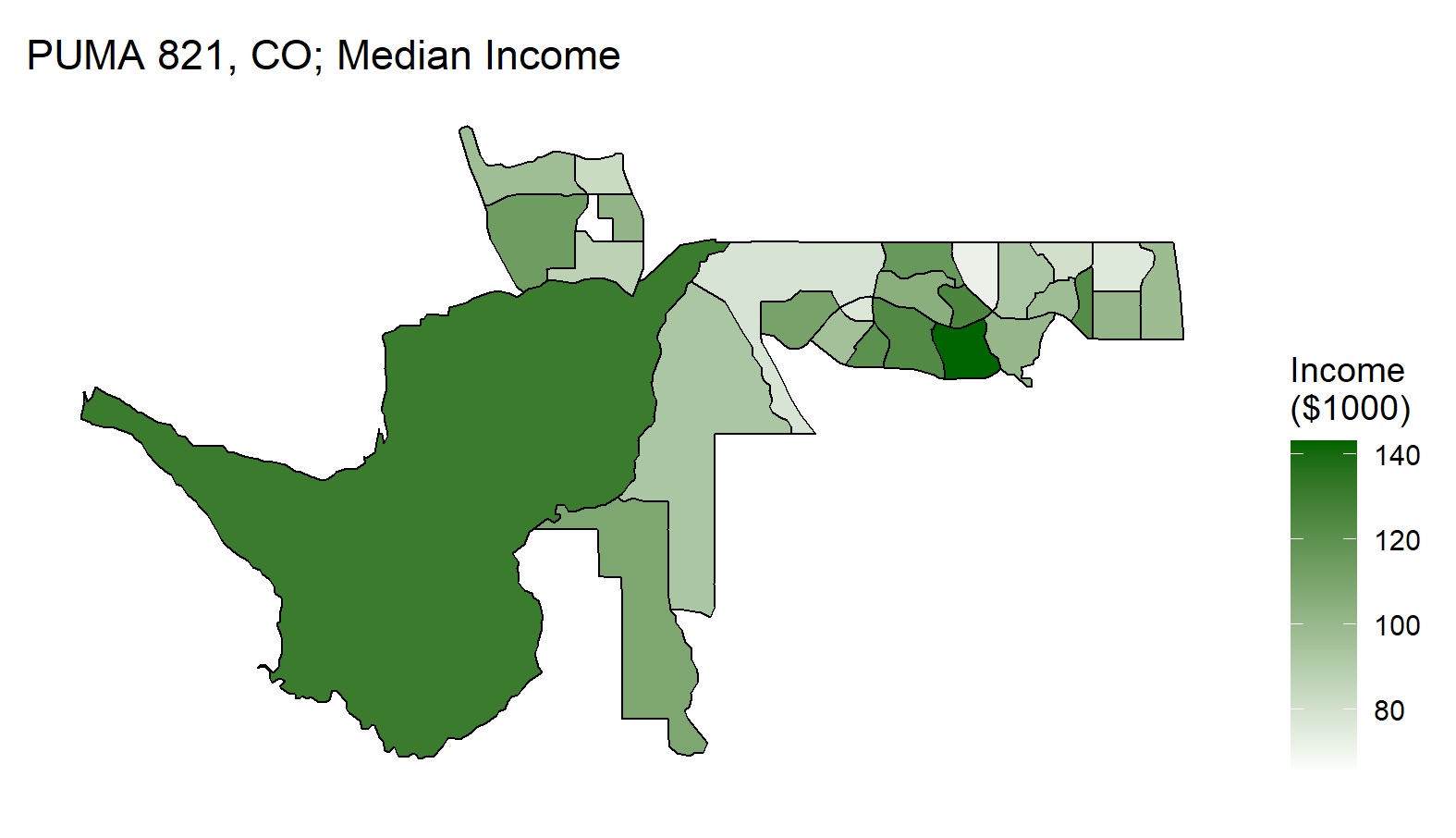}
  \includegraphics[width = 0.45\textwidth]{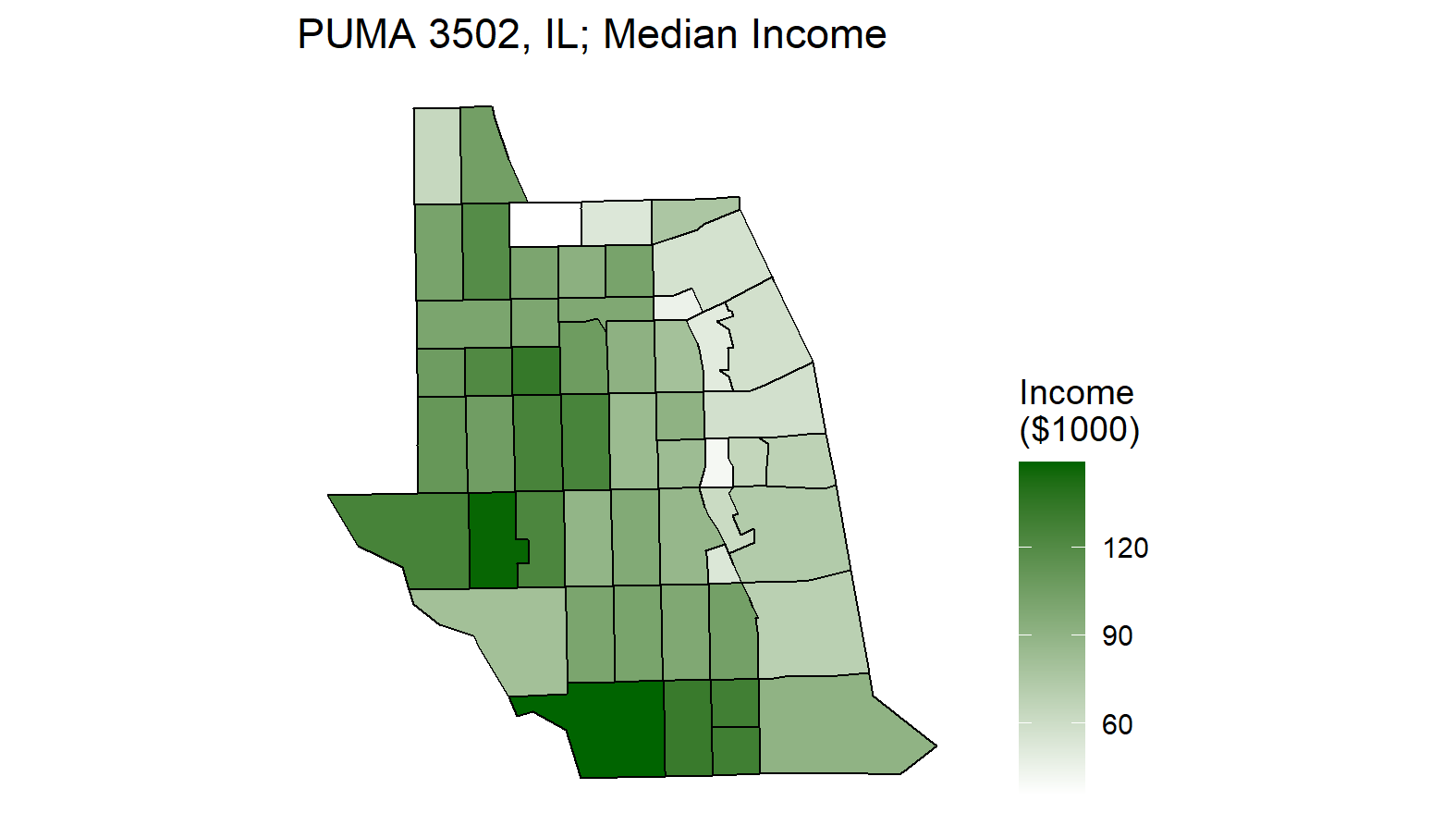}
  \includegraphics[width = 0.45\textwidth]{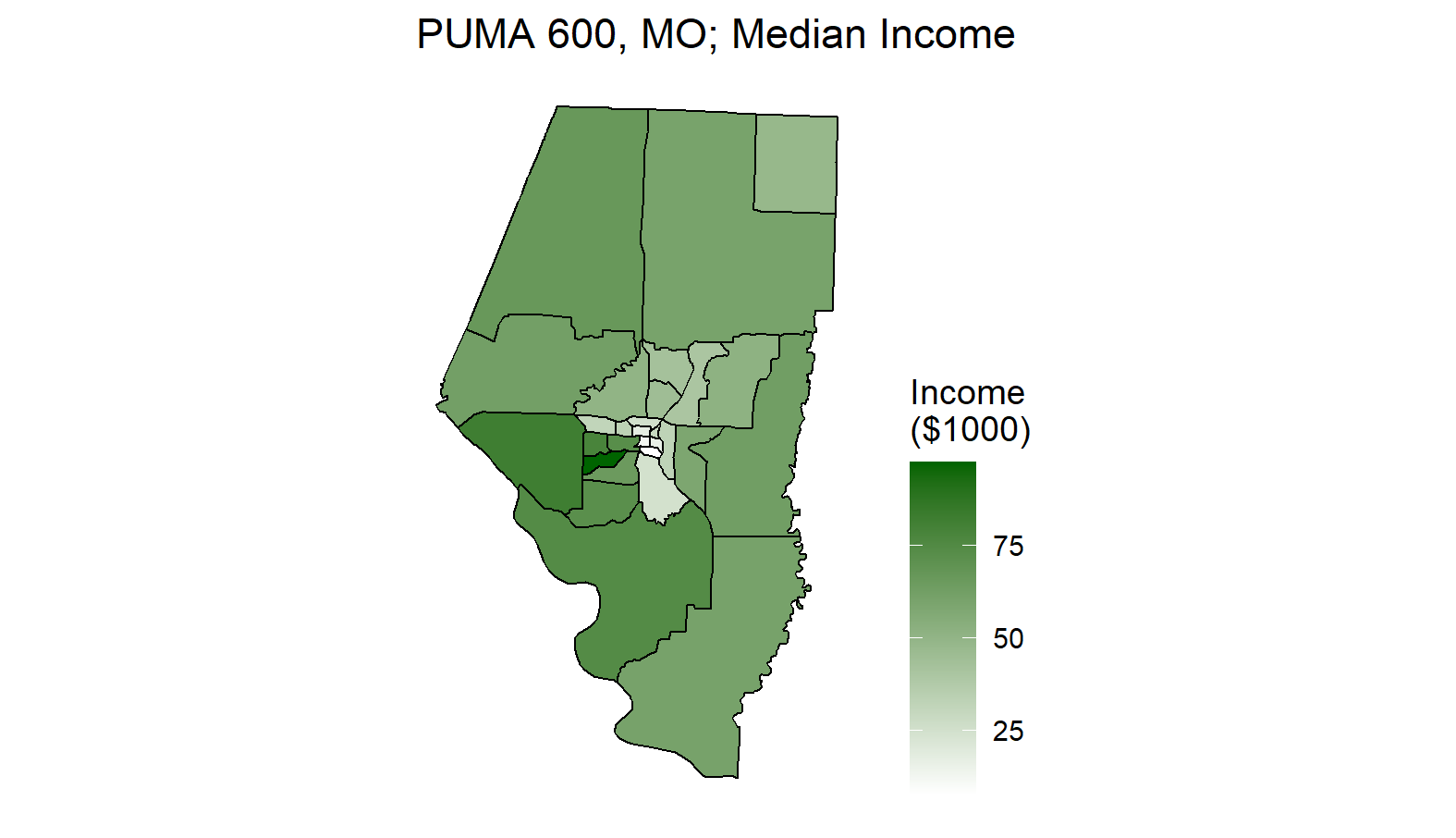}
  \includegraphics[width = 0.45\textwidth]{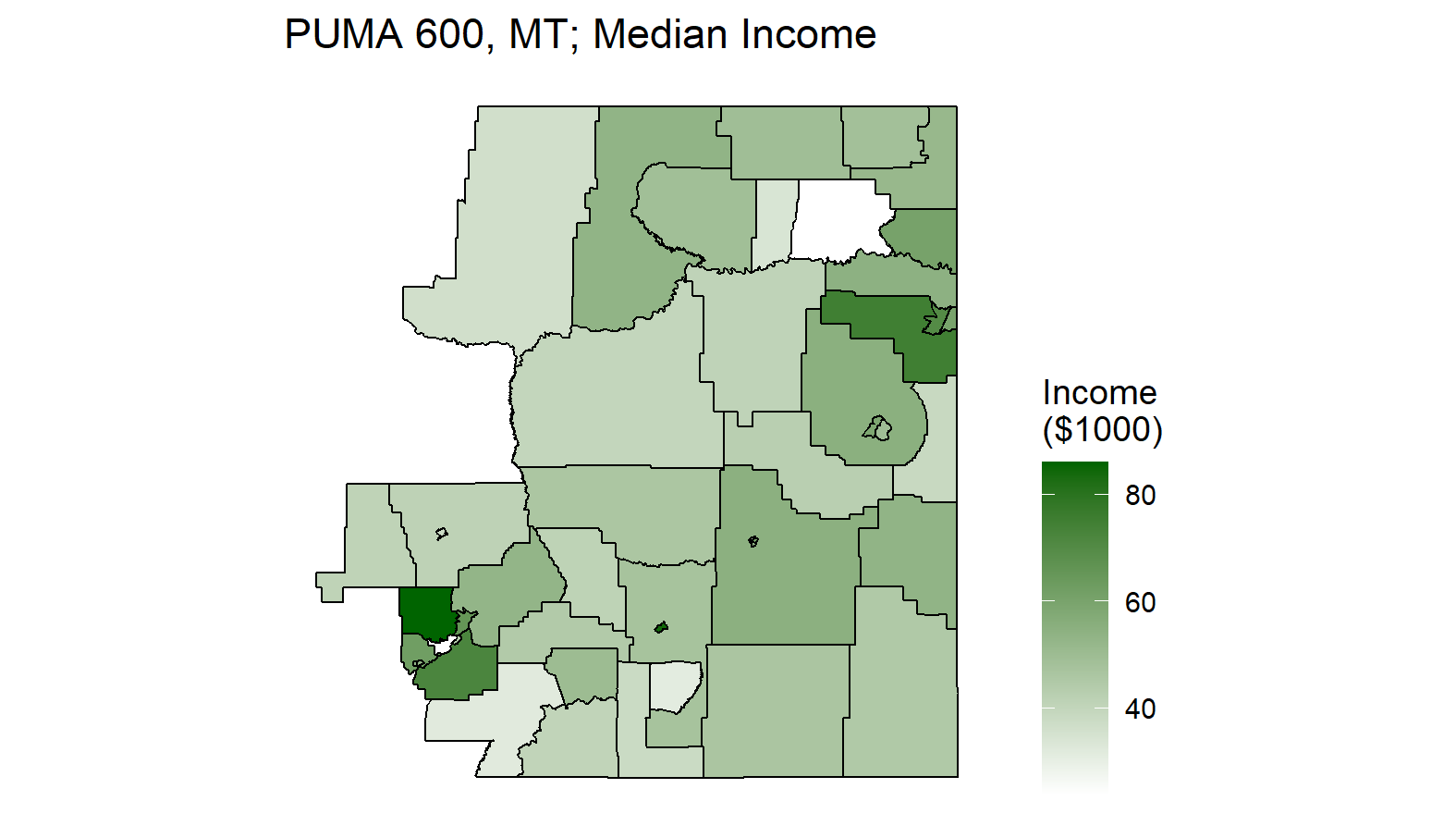}
  \includegraphics[width = 0.45\textwidth]{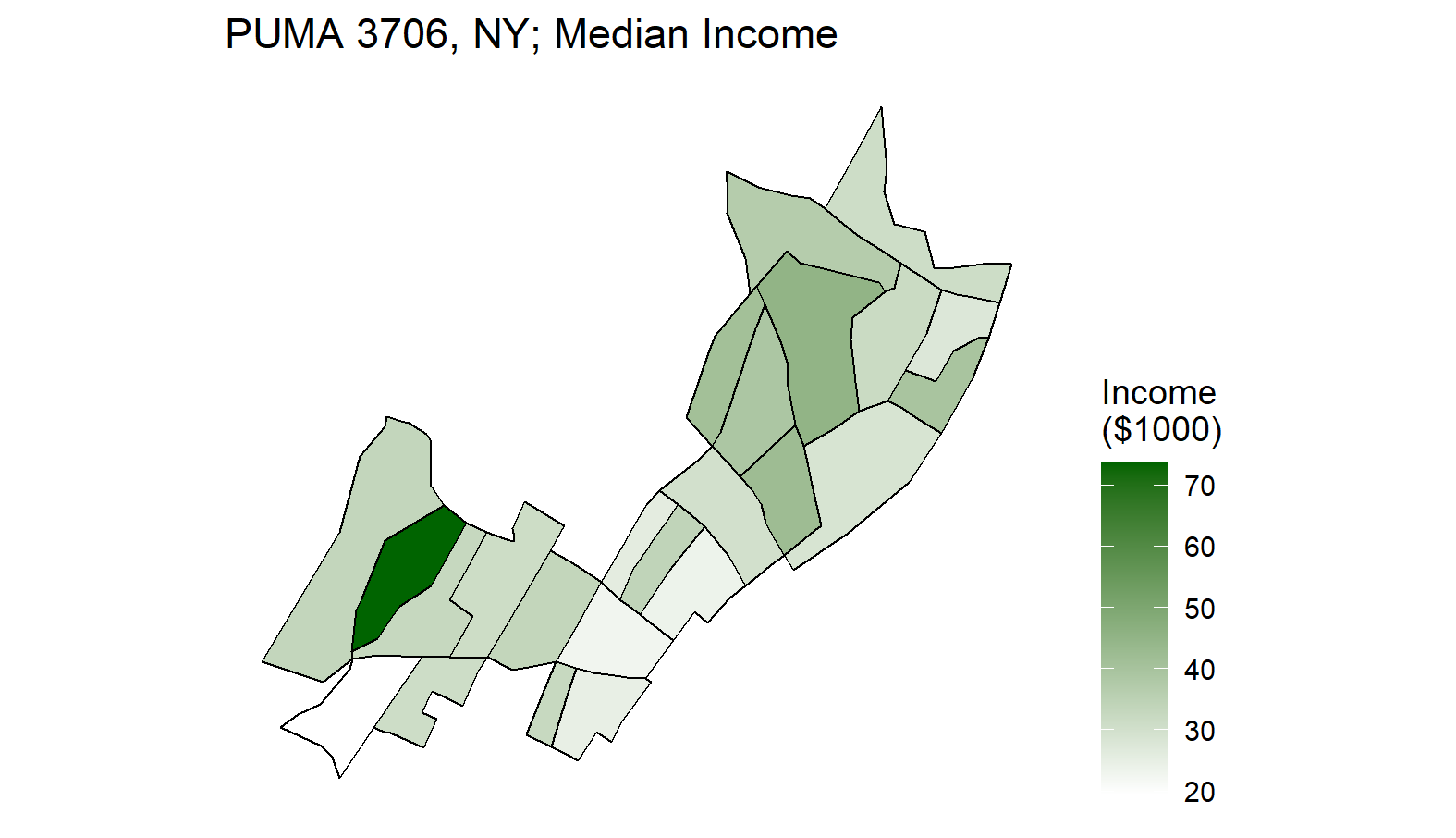}
  \caption{Maps of each PUMA used in the main text with each of the census tracts. Each tract within each PUMA is shaded according to the 2015 ACS 5-year period estimate of median household income.}
  \label{fig:allpumas.median}
\end{figure}

\clearpage

\section{LATENT PRLN DENSITY FUNCTIONALS}\label{app:functionals}
\setcounter{table}{0}
The latent PRLN density is given by
\begin{align*}
\pi(x) = \sum_{k=1}^Kp_kf_k(x).
\end{align*}
Let $k^*$ denote the largest knot which is less than an available estimate of the median. Then
\begin{align*}
  f_k(x) = &\  \frac{1}{\kappa_{k+1} - \kappa_k} \times \mathbbm{1}(\kappa_k < x \leq \kappa_{k+1}) && \mbox{ if $k^* \leq k^*$ },\nonumber\\
  = &\  \frac{\alpha_k\kappa_k^{\alpha_k}x^{-\alpha_k - 1}}{1 - \left(\frac{\kappa_k}{\kappa_{k+1}}\right)^{\alpha_k}} \times \mathbbm{1}(\kappa_k < x \leq \kappa_{k+1}) && \mbox{ if $k^* < k < K$ },\nonumber\\
  = &\  \alpha_k\kappa_kx^{-\alpha_k - 1} \times \mathbbm{1}(\kappa_K < x) && \mbox{ if $k = K$ }.
\end{align*}
For the $k$th bin, let $\mu_k$ denote its mean, $\sigma^2_k$ denote its variance, $F_k(x)$ denote its CDF, and $F^{-1}_k$ denote its quantile function, and $I_k$ denote its integrated Lorenz curve. Then the following sections derive formulas for several functionals in terms of these basic building blocks. 
\subsection{Mean}
Let $\mu = \E_\pi[x]$. Then
\begin{align*}
\mu = \sum_{k=1}^Kp_k\mu_k.
\end{align*}
Note that this requires that each $\mu_k$ exist.
\subsection{Variance}
Let $\sigma^2 = \var_{\pi}[x]$. Then the conditional variance formula yields
\begin{align*}
  \sigma^2 = \sum_{k=1}^Kp_k\sigma_k^2 + \sum_{k=1}^Kp_k(\mu_k - \mu)^2.
\end{align*}
Note that this requires that each $\sigma_k^2$ exist.
\subsection{CDF}
Let $\Pi(x)$ denote the CDF corresponding to $\pi(x)$. Then
\begin{align*}
  \Pi(x) = \begin{cases}
    0, & \mbox{ if } x \leq \kappa_1\\
    F_1(x), & \mbox{ if } \kappa_1 < x \leq \kappa_2\\
    p_1 + p_2*F_2(x), & \mbox{ if } \kappa_2 < x \leq \kappa_3 \\
  \vdots  & \vdots  \\
    \sum_{k=1}^{j-1}p_k + p_jF_j(x), &\mbox{ if } \kappa_j < x \leq \kappa_{j+1}\\
   \vdots & \vdots  \\
    \sum_{k=1}^{K-1}p_k + p_KF_K(x), &\mbox{ if } \kappa_{K} < x  \leq \kappa_{K+1} \\
    1, & \mbox{ if } \kappa_{K+1} \leq x.
  \end{cases}
\end{align*}
Note that bin probabilities are given by a difference in the CDF evaluated at two knots, i.e.
\begin{align*}
p_k = \Pi(\kappa_k) - \Pi(\kappa_{k-1}).
\end{align*}
\subsection{Quantile function}
Let $\Pi^{-1}$ denote the quantile function associated with $\pi(x)$. Then
\begin{align*}
  \Pi^{-1}(\tau) = \begin{cases}
    F^{-1}_1\left(\frac{\tau}{p_1}\right), & \mbox{ if } 0 \leq \tau \leq p_1 \\
    F^{-1}_2\left(\frac{\tau - p_1}{p_2}\right), & \mbox{ if } p_1 < \tau \leq p_1 + p_2 \\
  \vdots  & \vdots \\
    F^{-1}_j\left(\frac{\tau - \sum_{k=1}^{j-1}p_k}{p_j}\right), & \mbox{ if } \sum_{k=1}^{j-1}p_{k} < \tau \leq \sum_{k=1}^jp_{k} \\
  \vdots  & \vdots \\
    F^{-1}_K\left(\frac{\tau - \sum_{k=1}^{K-1}p_k}{p_K}\right), & \mbox{ if } \sum_{k=1}^{K-1}p_{k} < \tau \leq 1.
  \end{cases}
\end{align*}
Note that $\Pi^{-1}$ is not everywhere differentiable as a function of the $p_k$s.

\subsection{Integrated Lorenz curve}
The Lorenz curve for a PDF $f$ with associated CDF $F$ and mean $\mu$ is defined as
\begin{align*}
  L(\tau) = \frac{1}{\mu}\int_{-\infty}^{F^{-1}(\tau)}y f(y) dy.
\end{align*}
The integrated Lorenz curve is given by
\begin{align*}
I = \int_0^1L(\tau)d\tau = \int_{-\infty}^{\infty}L[F(x)]f(x)dx.
\end{align*}
Let $L_k$ denote the Lorenz curve for the $k$th bin. Let $\kappa_{k^*}$ denote the largest knot such that $\kappa_{k^*} \leq x = \Pi^{-1}(\tau)$ and note that for any Lorenz curve $L(0)=0$ and $L(1)=1$.  Then we can express the Lorenz curve for the latent PRLN density as
\begin{align*}
  L[\Pi(x)] & = \frac{1}{\mu}\int_{-\infty}^x y \pi(y) dy \\
          & = \frac{1}{\mu}\left[\sum_{j=1}^{k^*-1}p_j\mu_j + p_{k^*}\int_{\kappa_{k^*}}^xyf_{k^*}(y)dy\right] \\
          & = \frac{1}{\mu}\left[\sum_{j=1}^{k^*-1}p_j\mu_j + p_{k^*}\mu_{k^*}L_{k^*}[F_{k^*}(x)]\right] \\
          & = \frac{1}{\mu}\sum_{k=1}^Kp_k\mu_kL_k[F_k(x)].
\end{align*}
This implies that if we state the Lorenz curve in its original form as a pure function of $\tau$, we have
\begin{align*}
L(\tau) = \frac{1}{\mu}\sum_{k=1}^Kp_k\mu_kL_k\{F_k[F^{-1}(\tau)]\}.
\end{align*}
Then the integrated Lorenz curve can be written as
\begin{align*}
  I & = \frac{1}{\mu}\sum_{k=1}^Kp_k\mu_k\int_{-\infty}^{\infty}L_k[F_k(x)]\pi(x)dx \\
  & = \frac{1}{\mu}\sum_{k=1}^Kp_k\mu_k\sum_{j=1}^Kp_j\int_{\kappa_j}^{\kappa_{j+1}}L_k[F_k(x)]f_j(x)dx \\
    & = \frac{1}{\mu}\sum_{k=1}^Kp_k\mu_k\left[p_k\int_{\kappa_{k}}^{\kappa_{k+1}}L_k[F_k(x)]f_k(x)dx + \sum_{j=k+1}^Kp_j\right] \\
    & = \frac{1}{\mu}\sum_{k=1}^{K}p_k\mu_k\left[p_kI_k + \sum_{j=k+1}^Kp_j\right].
\end{align*}

\subsection{Distribution shares}
The Lorenz curve represents the proportion of aggregate income that goes to the lower $p$ proportion of the income distribution, for any $0 < p < 1$. So for an income distribution, distribution shares (income shares) are given by differences in the Lorenz curve evaluated at two points. For $0\leq \tau_1 < \tau_2 \leq1$ the aggregate income that goes to the distribution between $\tau_1$ and $\tau_2$ is given by
\begin{align*}
s(\tau_1, \tau_2) = L(\tau_2) - L(\tau_1).
\end{align*}
\subsection{Gini index}
The Gini index for a continuous distribution can be expressed in terms of the Lorenz curve as
\begin{align*}
G = 1 - 2\int_0^1L(\tau)d\tau = 1 - 2I
\end{align*}

\subsection{Applying to the latent PRLN density}
To apply this to the latent PRLN density, we need to find all of the building blocks for each bin type: uniform, truncated Pareto, and Pareto distributed.

\subsubsection{Uniform bins}
For uniform bins:
\begin{align*}
  \mu_k & = \frac{1}{2}(\kappa_{k+1} + \kappa_k) &&&\\
  \sigma_k^2 &= \frac{1}{12}(\kappa_{k+1} - \kappa_k)^2 &&&\\
  F_k(x) &= \frac{x - \kappa_k}{\kappa_{k+1} - \kappa_k} &&& \mbox{ for } \kappa_k \leq x \leq \kappa_{k+1} \\
  F^{-1}_k(\tau) & = \kappa_k + \tau (\kappa_{k + 1} - \kappa_k). &&&
\end{align*}
Then the Lorenz curve is
\begin{align*}
  L_k[F_k(x)] &= \frac{1}{\mu_k}\int_{\kappa_k}^x y f_k(y) dy \\
              &= \frac{1}{2\mu_k}\frac{x^2 - \kappa_k^2}{\kappa_{k+1} - \kappa_k}.
\end{align*}
The integrated Lorenz curve is
\begin{align*}
  I_k & = \int_{\kappa_k}^{\kappa_{k+1}}\frac{1}{2\mu_k}\frac{x^2 - \kappa_k^2}{(\kappa_{k+1} - \kappa_k)^2}dx \\
      & = \frac{1}{2\mu_k(\kappa_{k+1} - \kappa_k)^2}\left[\frac{\kappa_{k+1}^3}{3} - \frac{\kappa_{k}^3}{3} - (\kappa_{k+1} - \kappa_k)\kappa_k^2\right]\\
      & = \frac{1}{\mu_k}\left[\frac{(\kappa_{k+1} - \kappa_k)(\kappa_{k+1}^2 + \kappa_{k+1}\kappa_k + \kappa_k^2)}{6(\kappa_{k+1} - \kappa_k)^2} - \frac{\kappa_{k}^2}{2(\kappa_{k+1}-\kappa_k)}\right] \\
      & = \frac{1}{\mu_k}\frac{\kappa_{k+1}^2 + \kappa_{k+1}\kappa_{k} + \kappa_k^2 - 3 \kappa_k^2}{6(\kappa_{k+1} - \kappa_k)} \\
      & = \frac{\kappa_{k+1} + 2\kappa_k}{6\mu_k}\\
      & = \frac{\kappa_{k+1} + 2\kappa_k}{3(\kappa_{k+1} + \kappa_k)}\\
      & = \frac{1}{3}\left(1 + \frac{\kappa_k}{\kappa_k + \kappa_{k+1}}\right).
\end{align*}

\subsubsection{Pareto bins}
For Pareto distrubted bins
\begin{align*}
  \mu_K & = \frac{\alpha_K\kappa_K}{\alpha_K - 1} &&& \mbox{ if } \alpha_K > 1\\
  \sigma_K^2 &= \frac{\kappa_K^2\alpha_K}{(\alpha_K-1)^2(\alpha_K-2)} &&& \mbox{ if } \alpha_K > 2\\
  F_K(x) &= 1 - \left(\frac{\kappa_K}{x}\right)^{\alpha_K} &&& \mbox{ for } \kappa_K \leq x \\
  F^{-1}_K(\tau) & = \frac{\kappa_K}{(1 - \tau)^{1/\alpha_K}}. &&&
\end{align*}
Then the Lorenz curve is given by
\begin{align*}
  L_K[F_K(x)] & = \frac{1}{\mu_K}\int_{\kappa_K}^{x} \alpha_K\kappa_K^{\alpha_K} y^{-\alpha_K} dy \\
              & = \frac{\alpha_K\kappa_K^{\alpha_K}}{\mu_K}\left(-\frac{1}{\alpha_K - 1}y^{-\alpha_K + 1}\right)_{\kappa_K}^{x} \\
              & = \frac{1}{\mu_K}\frac{\alpha_K}{\alpha_K - 1}\left(\kappa_K - \kappa_K^{\alpha_K}x^{1 - \alpha_K}\right).
\end{align*}
Then the integrated Lorenz curve is
\begin{align*}
  I_K & = \int_{\kappa_K}^{\infty}L_K[F_K(x)]f_K(x) dx\\
      & = \frac{1}{\mu_k}\frac{\alpha_K}{\alpha_K - 1}\left(\kappa_K - \kappa_{K}^{\alpha_K}\int_{\kappa_K}^{\infty}x^{1 - \alpha_K}\alpha_K\kappa_K^{\alpha_K}x^{-\alpha_K - 1} dx\right) \\
        & = \frac{1}{\mu_k}\frac{\alpha_K}{\alpha_K - 1}\left(\kappa_K - \frac{1}{2}\int_{\kappa_K}^{\infty}2\alpha_K\kappa_K^{2\alpha_K}x^{-2\alpha_K} dx\right) \\
      & = \frac{1}{\mu_K}\frac{\alpha_K}{\alpha_K - 1}\left(\kappa_K - \frac{1}{2}\frac{2\alpha_K}{2\alpha_K - 1}\kappa_K\right)\\
      & = 1 - \frac{\alpha_K}{2\alpha_K - 1}.
\end{align*}

\subsubsection{Truncated Pareto bins}
For truncated Pareto bins
\begin{align*}
  \mu_k & = \frac{\alpha_k\kappa_k}{\alpha_k - 1}\frac{1 - \left(\frac{\kappa_k}{\kappa_{k+1}}\right)^{\alpha_k - 1}}{1 - \left(\frac{\kappa_k}{\kappa_{k+1}}\right)^{\alpha_k}} &&& \mbox{ if } \alpha_k > 1\\
  \sigma_k^2 &= \frac{\alpha_k}{\alpha_k - 2}\kappa_k^2\frac{1 - \left(\frac{\kappa_k}{\kappa_{k+1}}\right)^{\alpha_k - 2}}{1 - \left(\frac{\kappa_k}{\kappa_{k+1}}\right)^{\alpha_k}} - \frac{\alpha_k^2}{(\alpha_k-1)^2}\kappa_k^2\frac{\left[1 - \left(\frac{\kappa_k}{\kappa_{k+1}}\right)^{\alpha_k - 1}\right]^2}{\left[1 - \left(\frac{\kappa_k}{\kappa_{k+1}}\right)^{\alpha_k }\right]^2} &&& \mbox{ if } \alpha_k > 2\\
  F_k(x) &= \frac{1 - \left(\frac{\kappa_k}{x}\right)^{\alpha_k}}{1 - \left(\frac{\kappa_k}{\kappa_{k+1}}\right)^{\alpha_k}} &&& \mbox{ for } \kappa_k \leq x \leq \kappa_{k+1} \\
  F^{-1}_k(\tau) & = \frac{\kappa_k}{\left\{1 - \tau\left[1 - \left(\frac{\kappa_k}{\kappa_{k+1}}\right)^{\alpha_k}\right]\right\}^{1/\alpha_k}}. &&&
\end{align*}
The formulas for means and variances can be extended to $\alpha_k > 0$ so long as care is taken to account for special cases when $\alpha_k = 1$ and $\alpha_k = 2$.

Next, the Lorenz curve is given by
\begin{align*}
  L_k[F_k(x)] & = \frac{1}{\mu_k}\int_{\kappa_k}^{x} \frac{\alpha_k\kappa_k^{\alpha_k} y^{-\alpha_k}}{1 - \left(\frac{\kappa_k}{\kappa_{k+1}}\right)^{\alpha_k}} dy  \\
              & = \frac{1}{\mu_k}\frac{\alpha_k\kappa_k^{\alpha_k}}{1 - \left(\frac{\kappa_k}{\kappa_{k+1}}\right)^{\alpha_k}}\left[-\frac{1}{\alpha_k - 1}(x^{1 - \alpha_k} - \kappa_k^{1 - \alpha_k})\right] \\
              & = \frac{1}{\mu_k}\frac{\alpha_k}{\alpha_k - 1}\frac{\kappa_k^{\alpha_k}}{1 - \left(\frac{\kappa_k}{\kappa_{k+1}}\right)^{\alpha_k}}\left[\kappa_k^{1-\alpha_k} - x^{1 - \alpha_k}\right]\\
              & = \frac{\kappa_k}{\mu_k}\frac{\alpha_k}{\alpha_k - 1}\frac{1 - \left(\frac{\kappa_k}{x}\right)^{\alpha_k - 1}}{1 - \left(\frac{\kappa_k}{\kappa_{k+1}}\right)^{\alpha_k}}.
\end{align*}
Then the integrated Lorenz curve is given by
\begin{align*}
  I_k & = \int_{\kappa_k}^{\kappa_{k+1}}L_k[F_k(x)]f_k(x)dx \\
      & = \frac{1}{\mu_k}\frac{\alpha_k}{\alpha_k - 1}\frac{\kappa_k^{\alpha_k}}{1 - \left(\frac{\kappa_k}{\kappa_{k+1}}\right)^{\alpha_k}}\int_{\kappa_k}^{\kappa_{k+1}}\left[\kappa_k^{1-\alpha_k} - x^{1 - \alpha_k}\right]\frac{\alpha_k\kappa_k^{\alpha_k}}{1 - \left(\frac{\kappa_k}{\kappa_{k+1}}\right)^{\alpha_k}}x^{-\alpha_k - 1}dx \\
      & = \frac{1}{\mu_k}\frac{\alpha_k}{\alpha_k - 1}\frac{1}{1 - \left(\frac{\kappa_k}{\kappa_{k+1}}\right)^{\alpha_k}}\left[\kappa_k - \frac{1}{2}\frac{1 - \left(\frac{\kappa_k}{\kappa_{k+1}}\right)^{2\alpha_k}}{1 - \left(\frac{\kappa_k}{\kappa_{k+1}}\right)^{\alpha_k}}\int_{\kappa_k}^{\kappa_{k+1}}\frac{2\alpha_k\kappa_k^{2\alpha_k}}{1 - \left(\frac{\kappa_k}{\kappa_{k+1}}\right)^{2\alpha_k}}x^{-2\alpha_k}dx\right] \\
      & = \frac{1}{\mu_k}\frac{\alpha_k}{\alpha_k - 1}\frac{1}{1 - \left(\frac{\kappa_k}{\kappa_{k+1}}\right)^{\alpha_k}}\left[\kappa_k - \frac{1}{2}\frac{1 - \left(\frac{\kappa_k}{\kappa_{k+1}}\right)^{2\alpha_k}}{1 - \left(\frac{\kappa_k}{\kappa_{k+1}}\right)^{\alpha_k}}\frac{2\alpha_k}{2\alpha_k - 1}\kappa_k\frac{1 - \left(\frac{\kappa_k}{\kappa_{k+1}}\right)^{2\alpha_k - 1}}{1 - \left(\frac{\kappa_k}{\kappa_{k+1}}\right)^{2\alpha_k}}\right] \\
      & = \frac{\kappa_k}{\mu_k}\frac{\alpha_k}{\alpha_k - 1}\frac{1}{1 - \left(\frac{\kappa_k}{\kappa_{k+1}}\right)^{\alpha_k}}\left[1 - \frac{1 - \left(\frac{\kappa_k}{\kappa_{k+1}}\right)^{2\alpha_k - 1}}{1 - \left(\frac{\kappa_k}{\kappa_{k+1}}\right)^{\alpha_k}}\frac{\alpha_k}{2\alpha_k - 1}\right] \\
      & = 1 - \frac{\alpha_k}{2\alpha_k - 1}\frac{1 - \left(\frac{\kappa_k}{\kappa_{k+1}}\right)^{2\alpha_k - 1}}{1 - \left(\frac{\kappa_k}{\kappa_{k+1}}\right)^{\alpha_k}}\\
      & = 1 - \frac{\alpha_k\kappa_{k+1}^{1 - \alpha_k}}{2\alpha_k - 1}\frac{\kappa_{k+1}^{2\alpha_k - 1} - \kappa_k^{2\alpha_k - 1}}{\kappa_{k+1}^{\alpha_k} - \kappa_k^{\alpha_k}}.
\end{align*}

\section{GENERATING THE SYNTHETIC POPULATION}\label{app:synpop}
\setcounter{table}{0}
We construct the population in our simulation study to have the same number of households per tract as the 2014 ACS 5-year period estimates of household population for the Boone County, MO PUMA. We then divide the population into the same 106 strata that exist in the 2014 Boone County 5-year PUMS -- a stratum is defined as all observations with the same survey weight. The population of each stratum is assumed to be to $n_sw_s$ where $n_s$ is the sample size of stratum $s$ in the PUMS, and $w_s$ is the survey weight associated with stratum $s$. To fully specify the population we need to know number of households in each tract/stratum combination, though in reality this is unknown. Nevertheless, we know that the PUMS strata are based in part on census tracts \citep{acs2015}, so in our synthetic population we assign the households in a given stratum to a small number of tracts using an algorithm that produces tract and stratum assignments that are closely related.

Next, an income is generated for each household using a two-component mixture of lognormals with parameters that depend on both their tract and stratum. The resulting tract-level distributions are mixtures of lognormals. Figure~\ref{fig:pops} in the Supplementary Materials contains maps of the true tract-level means, medians, and standard deviations of income for the synthetic population.

The above description omits two important pieces of how the population is generated. First, how strata are assigned to tracts, and second, how incomes are generated for each tract/stratum combination. We take these in turn.

\subsection{Assigning strata to tracts}
Algorithm~\ref{alg:stratum2tract} describes how strata are assigned to tracts. Essentially, for each tract, we randomly select a stratum, then assign as much of that stratum as we can to the tract. If the stratum fully fits in the tract (along with the strata already assigned to it), then the stratum is deleted from the pool of available strata, and a new one is randomly selected to repeat the process. If the stratum does not fit, then the stratum is returned to the pool of available strata with its remaining population, and we move on to the next tract.
\begin{algorithm}[ht]
  \begin{algorithmic}
  \item
    \FORALL{tract}
    \STATE{Initialize tract.pop = 0}
    \WHILE{tract.pop $<$ tract.popest}
    \STATE{Randomly select a stratum with stratum.pop $>$ 0}
    \STATE{Set P $=$ MIN(stratum.pop, tract.popest - tract.pop)}
    \STATE{Assign P  members of the stratum to tract}
    \STATE{Set target.pop $+=$ P}
    \STATE{Set stratum.pop $-=$ P}
    \ENDWHILE
    \ENDFOR
  \end{algorithmic}
  \caption{Assign strata to tracts. Assume that tract.popest is the desired population of the tract, and that stratum.pop is initialized with the assigned population of the stratum.}
  \label{alg:stratum2tract}
\end{algorithm}

\subsection{Generating incomes for tract/stratum combinations}
Generating the incomes is more complex. For each tract/stratum combination we define a two-component mixture of lognormal distributions, using the PUMS data as a guide. To do this, we need several intermediate quantities. First, using the PUMS data, let $\widehat{m}$ denote the sample mean of $z = \log(\mathrm{income} + 1)$ and let $\widehat{s}$ denote the sample standard deviation. We use the offset of one because there are incomes equal to zero in the dataset.

Next for each stratum, we compute the a measure of dispersion of $z$ and a measure of how far $z$ tends to be away from the the PUMA mean. Let $i=1,2,\dots,n_s$ index observations in stratum $s$, and $z_{is}$ denote log offset income for each of those observations, as in the previous paragraph. Then define
\begin{align*}
D_s = \frac{1}{n_s + 5}\sum_{i=1}^{n_s}(z_{is} - \widehat{m}).
\end{align*}
This is a measure of how far the stratum tends to be from the PUMA average, regularized toward zero since many strata have as few as one observation. Similarly, define
\begin{align*}
H_s^2 = \frac{n_s}{n_s + 500}\frac{1}{n_s}\sum_{i=1}^{n_s}(z_{is} - \overline{z}_s)^2 + \frac{500}{n_s + 500}\widehat{s}^2,
\end{align*}
where $\overline{z}_s$ is the mean of $z_{is}$ in stratum $s$. This is a measure of dispersion in the stratum, again regularized to be much closer the PUMA level dispersion. Note that we divide by $n_s$ instead $n_s - 1$ to avoid dividing by zero in strata with only one member.

Finally, we need a tract-level and a stratum-level covariate to use these quantities with. For a tract $r$, let $\mathrm{dist}_r$ denote the average distance of tract $r$ from the center of the bounding box containing the PUMA, and let $ \mathrm{sdist}_r = (\mathrm{dist}_r - \mathrm{mean}(\mathrm{dist}_{1:R}))/\mathrm{sd}(\mathrm{dist}_{1:R})$ denote the scaled distance from the center for $r$. Next let $w_s$ denote the unique weight associated with stratum $s$. Finally let $W_s = (\log w_s - \mathrm{mean}(\log w_{1:S})) / \mathrm{sd}(\log w_{1:S})$ denote the scaled log weight for $s$.

Using these quantities, we need to choose the mean parameters $\mu_1$ and $\mu_2$, the standard deviation parameters $\sigma_1$ and $\sigma_2$, and the mixture weight $\omega$, all for a given tract/stratum combination $(r,s)$. We use the following quantities:
\begin{align*}
  \omega &= \frac{1}{1 + \exp[0.2 \mathrm{sdist}_r + 0.2 * W_s]}\\
  \mu_1 &= 0.87\widehat{m} - 0.3 \mathrm{sdist}_r + D_s\\
  \mu_2 &= 1.05\widehat{m} - 0.2 \mathrm{sdist}_r + 1.5 D_s\\
  \sigma_1 & = \exp\left[\frac{\mathrm{sdist}_r}{5} - \frac{\log H_s}{5}\right]\\
  \sigma_2 & = 0.6\exp\left[\frac{\mathrm{sdist}_r}{5} - \frac{\log H_s - \log 0.6}{5}\right].
\end{align*}
We arrived at these settings through exploratory analysis until we found a population of incomes that looked somewhat like a real income distribution. The distribution includes natural spatial variation across tracts and variation across strata, in an attempt to mimic the observed data.

\clearpage

\section{EVALUATING POINT ESTIMATES}\label{app:acstab}
\setcounter{table}{0}
\begin{table}[ht]
  \centering
  \begin{tabular}{lllrrrrrr}
    \hline
          & Estimator & 20th & 40th & 60th & 80th & 95th & Gini \\
    \hline
    MAD   & PRLN         & 1906 & 2014 & 2093 & 4417 & 7369 & 0.0176 \\
          & Mean         & 2207 & 1926 & 2020 & 5510 & 8114 & 0.0123 \\
          & Median       & 2200 & 1915 & 1981 & 5641 & 9172 & 0.0122 \\\hline
    MAPE  & PRLN         & 3.44 & 2.36 & 1.81 & 2.78 & 3.38 & 4.60 \\
          & Mean         & 4.18 & 2.35 & 1.72 & 3.28 & 3.79 & 3.48 \\
          & Median       & 4.20 & 2.35 & 1.69 & 3.36 & 4.39 & 3.39 \\\hline
    RMSE  & PRLN         & 2203 & 2614 & 2813 & 5584 & 9998  & 0.0251 \\
          & Mean         & 2591 & 2332 & 2629 & 7371 & 10550 & 0.0149 \\
          & Median       & 2599 & 2358 & 2591 & 7493 & 11891 & 0.0152 \\\hline
    RMSPE & PRLN         & 3.86 & 3.00 & 2.39 & 3.53 & 4.40 & 6.25 \\
          & Mean         & 4.89 & 2.83 & 2.28 & 4.06 & 4.91 & 4.17 \\
          & Median       & 4.96 & 2.88 & 2.22 & 4.17 & 5.67 & 4.07 \\
    \hline
  \end{tabular}
  \caption{MAD, MAPE, RMSE, and RMSPE for several estimates of the held out quantiles and Gini coefficient for the CO PUMA. The estimates are the PRLN estimate (PRLN), and the posterior predictive mean and median from L-PRLN (Mean and Median).}
  \label{tab:co.metric}
\end{table}

\begin{table}[ht]
  \centering
  \begin{tabular}{lllrrrrrr}
    \hline
           & Estimator & 20th & 40th & 60th & 80th & 95th & Gini \\
    \hline
    MAD    & PRLN         & 1270  & 1705 & 3658 & 8423  & 36108 & 0.066 \\
           & Mean         & 2669  & 2429 & 3613 & 5740  & 15387 & 0.020 \\
           & Median       & 2544  & 2363 & 3793 & 5737  & 14153 & 0.022 \\
    \hline
    MAPE   & PRLN         & 3.12  & 2.57 & 3.13 & 4.39  & 16.00 & 13.12 \\
           & Mean         & 8.18  & 3.29 & 3.03 & 3.44  & 7.64  & 3.94 \\
           & Median       & 7.48  & 3.28 & 3.15 & 3.49  & 6.98  & 4.41 \\
    \hline
    RMSE   & PRLN         & 1870  & 2474 & 4927 & 13429 & 48647 & 0.089 \\
           & Mean         & 3545  & 3119 & 5306 & 7212  & 18034 & 0.025 \\
           & Median       & 3354  & 3063 & 5634 & 7279  & 16479 & 0.028 \\
    \hline
    RMSPE  & PRLN         & 4.18  & 3.82 & 4.09 & 6.21  & 20.60 & 17.45 \\
           & Mean         & 11.63 & 4.08 & 4.08 & 4.19  & 9.04  & 4.76 \\
           & Median       & 10.62 & 4.17 & 4.25 & 4.38  & 8.20  & 5.36 \\
    \hline
  \end{tabular}
  \caption{MAD, MAPE, RMSE, and RMSPE for several estimates of the held out quantiles and Gini coefficient for the IL PUMA. The estimates are the PRLN estimate (PRLN), and the posterior predictive mean and median from L-PRLN (Mean and Median).}
  \label{tab:il.metric}
\end{table}

\begin{table}[ht]
  \centering
  \begin{tabular}{lllrrrrrr}
    \hline
            & Estimator & 20th & 40th & 60th & 80th & 95th & Gini \\
    \hline
    MAD     & PRLN         & 492   & 1053  & 2040 & 2981 & 7925  & 0.021 \\
            & Mean         & 1229  & 1467  & 2173 & 3471 & 10037 & 0.019 \\
            & Median       & 1128  & 1514  & 2087 & 3550 & 10062 & 0.020 \\
    \hline
    MAPE    & PRLN         & 2.96  & 2.80  & 3.35 & 3.83 & 5.14  & 4.30 \\
            & Mean         & 6.27  & 3.72  & 3.84 & 4.51 & 6.14  & 3.88 \\
            & Median       & 5.63  & 3.80  & 3.66 & 4.56 & 6.10  & 4.15 \\
    \hline
    RMSE    & PRLN         & 714   & 1561  & 2826 & 3867 & 11217 & 0.031 \\
            & Mean         & 1818  & 2019  & 2840 & 4318 & 14959 & 0.026 \\
            & Median       & 1609  & 2072  & 2807 & 4577 & 15089 & 0.028 \\
    \hline
    RMSPE   & PRLN         & 4.07  & 3.67  & 4.29 & 5.32 & 6.42  & 5.60 \\
            & Mean         & 8.28  & 4.66  & 4.86 & 5.89 & 7.90  & 4.82 \\
            & Median       & 7.23  & 4.76  & 5.00 & 6.37 & 7.90  & 5.13 \\
    \hline
  \end{tabular}
  \caption{MAD, MAPE, RMSE, and RMSPE for several estimates of the held out quantiles and Gini coefficient for the MO PUMA. The estimates are the PRLN estimate (PRLN), and the posterior predictive mean and median from L-PRLN (Mean and Median).}
  \label{tab:mo.metric}
\end{table}

\begin{table}[ht]
  \centering
  \begin{tabular}{lllrrrrrr}
    \hline
            & Estimator & 20th & 40th & 60th & 80th & 95th & Gini \\
    \hline
    MAD     & PRLN          & 542   & 1100 & 1581 & 2312 & 6362  & 0.015 \\
            & Mean          & 973   & 1450 & 1683 & 3161 & 7593  & 0.012 \\
            & Median        & 1015  & 1381 & 1725 & 3194 & 7561  & 0.013 \\
    MAPE    & PRLN          & 2.48  & 2.76 & 2.54 & 2.40 & 4.24  & 3.39 \\
            & Mean          & 4.63  & 3.71 & 2.67 & 3.34 & 4.91  & 2.71 \\
            & Median        & 4.76  & 3.52 & 2.72 & 3.36 & 4.79  & 2.95 \\
    RMSE    & PRLN          & 739   & 1424 & 2081 & 3318 & 8187  & 0.021 \\
            & Mean          & 1235  & 1869 & 2370 & 3893 & 9107  & 0.016 \\
            & Median        & 1282  & 1869 & 2490 & 3979 & 9627  & 0.018 \\
    RMSPE   & PRLN          & 3.37  & 3.50 & 3.26 & 3.31 & 5.48  & 4.63 \\
            & Mean          & 5.95  & 4.80 & 3.60 & 4.01 & 5.71  & 3.54 \\
            & Median        & 5.97  & 4.80 & 3.76 & 4.11 & 5.82  & 3.99 \\
    \hline
  \end{tabular}
  \caption{MAD, MAPE, RMSE, and RMSPE for several estimates of the held out quantiles and Gini coefficient for the MT PUMA. The estimates are the PRLN estimate (PRLN), and the posterior predictive mean and median from L-PRLN (Mean and Median).}
  \label{tab:mt.metric}
\end{table}

\begin{table}[ht]
  \centering
  \begin{tabular}{lllrrrrrr}
    \hline
            & Estimator & 20th & 40th & 60th & 80th & 95th & Gini \\
    \hline
    MAD     & PRLN          & 527   & 479  & 1687 & 2372 & 6118 & 0.022 \\
            & Mean          & 917   & 865  & 1850 & 2587 & 5698 & 0.022 \\
            & Median        & 831   & 654  & 1642 & 2458 & 6544 & 0.023 \\
    MAPE    & PRLN          & 3.96  & 1.86 & 3.87 & 3.38 & 5.55 & 4.42 \\
            & Mean          & 7.52  & 3.60 & 4.21 & 3.73 & 5.13 & 4.27 \\
            & Median        & 6.72  & 2.73 & 3.68 & 3.51 & 5.86 & 4.49 \\
    RMSE    & PRLN          & 709   & 658  & 2301 & 3132 & 7868 & 0.039 \\
            & Mean          & 1134  & 1050 & 2505 & 3208 & 7058 & 0.038 \\
            & Median        & 1015  & 912  & 2397 & 3094 & 7901 & 0.039 \\
    RMSPE   & PRLN          & 5.13  & 2.57 & 5.21 & 4.10 & 7.27 & 6.86 \\
            & Mean          & 9.28  & 4.40 & 5.39 & 4.33 & 6.28 & 6.64 \\
            & Median        & 8.20  & 3.89 & 5.06 & 4.10 & 6.98 & 6.96 \\
    \hline
  \end{tabular}
  \caption{MAD, MAPE, RMSE, and RMSPE for several estimates of the held out quantiles and Gini coefficient for the NY PUMA. The estimates are the PRLN estimate (PRLN), and the posterior predictive mean and median from L-PRLN (Mean and Median).}
  \label{tab:ny.metric}
\end{table}

\clearpage

\section{SEGREGATION INDEX EIV DATA}\label{app:data}
\setcounter{table}{0}
All of the data used in the income segregation index regressions was sourced from the ACS. We attempted to construct each variable used by \citet{reardon2011income} in their Table~4. For this exercise, we obtained metro-level 5-year ACS period estimates for a variety of variables, for each of the top 100 metro areas by population according to the 2018 5-year ACS estimates.

For many variables, we had to transform the ACS estimate in order to get it into the form needed for the model. MOEs and SEs for these ``derived estimates'' were derived using Census guidelines in the 2018 ACS general handbook \citep{us2018understanding}. Note that these MOEs and SEs are approximations, especially to the extent that they do not take into account correlation between the errors of multiple input estimates to a derived estimate. Below are tables describing the various pieces of ACS data needed for this exercise.

The only metro-level variables used by \citet{reardon2011income} that we could not construct a reasonable analogue for were percent of families with female householder by race and household income Gini index by race. The former was omitted from the analysis, and we used L-PRLN to estimate the latter, described in Appendix~\ref{app:gini}. Any metro area which does not have all necessary estimates available for a given regression is omitted from that regression. Additionally, if fewer than five census tracts from a metro area were available to compute the information theory and divergence indices for a given group of households, then that metro area was omitted from all regressions for that household group. As a result, $N=83$ in the all households and white households regressions, and $N=79$ in the black households regressions.

\begin{table}[ht]
  \centering
  \scriptsize{
  \begin{tabular}{|m{9em}|m{3cm}|m{9cm}|}
    \hline
    Their Variable & ACS Variable & Notes \\
    \hline
    Unemployment Rate & S2301\_C04\_001 & Also have labor force participation rate (S2301\_C02\_001) and employment to population ratio (S2301\_C03\_001).\\\hline

    Percent below age 18 & NA & Constructed via 100 minus S0101\_C02\_026, which is percent 18 and up. MOE is the same.\\\hline
    Percent age 65 \& up & S0101\_C02\_030 & \\\hline
    Percent of age 25 and up with at least a HS degree & S1501\_C02\_014 & \\\hline
    Per capita income & B19301\_001 & \\\hline
    Percent foreign born & DP02\_0092P & \\\hline
    Percent employed in manufacturing & NA & Constructed from total employed civilian population age 16 and older (S2405\_C01\_001) and total in manufacturing (S2405\_C01\_004) \\\hline
    Percent employed in construction & NA & Constructed from total employed civilian population age 16 and older (S2405\_C01\_001) and total in construction (S2405\_C01\_003) \\\hline
    Percent employed in FIRE (finance, insurance, and real estate) & NA & Constructed from total employed civilian population age 16 and older (S2405\_C01\_001) and total in FIRE (S2405\_C01\_009) \\\hline
    Percent employed in professional / managerial (information, FIRE, education, health, other prof, public admin) & NA & Constructed from total employed civilian population age 16 and older (S2405\_C01\_001), total in information (S2405\_C01\_008), total in FIRE (S2405\_C01\_009), total in education and health (S2405\_C01\_011), total in other professional (S2405\_C01\_010), and total in public admin (S2405\_C01\_014) \\\hline
    Percent of families with female householder & NA & Constructed from total families with male householder (B09019\_005) and total families with female householder (B09019\_006)\\\hline
    Percent of population in the same house as five years ago & NA & ACS does not provide this information. However, instead we constructed percent of population in the same house as one year ago with total population (B07204\_001) and total population in the same house one year ago (B07204\_002)\\\hline
    Percent of population in a different house from five years ago, but in the same county & NA & ACS does not provide this information. However, instead we construct a similar variable for one year ago with total population (B07204\_001), total population in a different house in the same town and in the same county (B07204\_005), and total population in a different house in a different town in the same county (B07204\_008)\\\hline
    Percent of housing that was built 1, 5, and 10 years ago & NA & ACS only provides percent of housing built within certain dates. Variable DP04\_0017P is the most recent set of dates, but it is inconsistent across years. For 2018 5-year estimates, it is the percent of housing built in 2014 or later, encompassing all 5 years of the period estimates. We used this variable.\\\hline
  \end{tabular}
}
\caption{Matching metro level variables with ACS variables}
\label{tab:metro.acs}
\end{table}

\begin{table}[ht]
  \centering
  \scriptsize{
  \begin{tabular}{|m{9em}|m{3cm}|m{9cm}|}
    \hline
    Their Variable & ACS Variable & Notes \\
    \hline
    Total Population by race & Table B02001 & \\\hline
    Unemployment Rate by race & S2301\_C04\_012 / S2301\_C04\_013 & \\\hline
    Percent of age 25 and up with at least a HS degree by race & S1501\_C02\_029 and S1501\_C02\_035 & \\\hline
    Per capita income & NA & Constructed from aggregate income tables B19025A and B19025B and population estimates, and SEs are adjusted accordingly.\\\hline
    Percent of families with female householder by race & NA & This variable could not be constructed from ACS estimates. \\\hline
    \end{tabular}
    }
    \caption{Matching metro level race based variables with ACS variables}
    \label{tab:race.acs}
\end{table}

\clearpage

\section{HOUSEHOLD LEVEL GINI INDEX ESTIMATION BY RACE}\label{app:gini}
\setcounter{table}{0}
To estimate household level Gini indices by race for each metro area, we employ L-PRLN. The available income estimates are described in Table~\ref{tab:race.income.ests}. Since metro areas have such large populations, we do not use the posterior predictive distribution to construct the Gini indices. Instead we directly use the formula for the Gini index of the latent PRLN density in Appendix~\ref{app:functionals} for every iteration in the posterior sample. Then we use the mean and standard deviation of the posterior sample of Gini indices for each metro area as the estimate and standard error in the EIV covariate matrix.

\begin{table}[ht]
  \centering
  \scriptsize{
  \begin{tabular}{|m{9em}|m{4cm}|m{8cm}|}
    \hline
    Income Variable & ACS Table & Notes \\
    \hline
    Household bin estimates & Table B19001A/B & The table is counts, so we convert to proportions and adjust the SE appropriately.\\\hline
    Household median income & Table B19013A/B &  \\\hline
    Household mean income & Table B19025A/B & The table is aggregate income, so we convert to mean income and adjust the SE appropriately. \\\hline
  \end{tabular}
  }
  \caption{Matching income variables to income variables by race}
  \label{tab:race.income.ests}
\end{table}

\clearpage

\section{COMPUTING SEGREGATION INDICES}\label{app:compute.indices}
\setcounter{table}{0}
Let $i=1,2,\dots,I$ index Census tracts in a metro area, each with population $N_i$, income CDF $F_i$ and corresponding PDF $f_i$. Let $w_i = N_i / \sum_{j=1}^IN_j$. Then the income CDF and PDF, respectively, for the entire metro area are given by \eqref{eq:app.metro.dist}
\begin{align}
  F(y) &= \sum_{i=1}^Iw_iF_i(y) &&& f(y) &= \sum_{i=1}^If_i(y). \label{eq:app.metro.dist}
\end{align}
The rank-order information theory index is given by \eqref{eq:app.hr}, while the divergent index is given by \eqref{eq:app:div}
\begin{align}
  H_R &= \sum_{i=1}^Iw_i\frac{E(F||F) - E(F||F_i)}{E(F||F)}, &&& E(G||F) &= \int_{-\infty}^{\infty}e[F(y)]dG(y) \label{eq:app.hr}\\
  D &= \sum_{i=1}^Iw_i D(f_i||f), &&& D(g||f)&= \int_{-\infty}^{\infty}\log \frac{g(y)}{f(y)}g(y)dy. \label{eq:app:div}
\end{align}
In both cases we approximate the integrals using importance sampling techniques.

Strictly speaking both $H_R$ and $D$ are functions of the model parameters for each census tract in the metro area. The upshot is that we need to approximate these integrals for each draw from the posterior distribution of the tract-level model parameters and obtain a joint posterior distribution of $D$, $H_R$, and both of their associated standard errors.

In a naive Monte Carlo approximation of $D$ given $M$ draws from the posterior, and $L$ draws for the Monte Carlo simulation, and $I$ tracts, characterizing the posterior of $D$ requires $\mathcal{O}(MLI^2)$ tract-level log density evaluations -- note that computing $f = \sum_{i=1}^Iw_if_i$ requires $I$ tract level log density evaluations, and $\mathcal{O}(MLI)$ simulations. This can be quite slow.

So we use two approaches to speed this up. First, since the latent PRLN density is piecewise defined, we break each integral into $K$ piecewise  sub-integrals. In income bins where all $I$ tracts are uniformly distributed, this allows us to solve the sub-integrals analytically. Second, for a given income bin, we only simulate one set of incomes that is used to compute the sub-intregrals for all tracts in that bin. This reduces the number of tract-level log density evaluations to $\mathcal{O}(MLI)$ and the number of simulations to $\mathcal{O}(ML)$, at the cost of inducing error correlations between each of the $D(f_i||f)$s. This correlation structure must be taken into account to compute the correct standard error for our estimate of $D$. We use the same basic approach for approximating $H_R$. Details follow in Appendix~\ref{app:compute.div}.

The computational problem for $H_R$ is easier, and our approach is simpler -- we use a straightforward importance sampler estimator, again using the same importance samples for all tracts within a metro area. But we do not separately sample from each bin. Details are in Appendix~\ref{app:compute.info}.

\subsection{DIVERGENCE INDEX}\label{app:compute.div}
Let $D_i = D(f_i||f)$, and suppose that $f_i$ is a latent PRLN density. Then we have
\begin{align*}
f_i(y) = \sum_{k=1}^Kp_{ik}f_{ik}(y)\mathbbm{1}(\kappa_{k} < y \leq \kappa_{k+1}),
\end{align*}
where $k$ indexes income bins, $p_{ik}$ is the probability the $i$th tract assigns to the $k$th bin, and $f_{iK}$ is the probability density of the $i$th tract in the $k$th bin. Then we can plug this into the formula for $D_i$ to obtain
\begin{align*}
  D_i & = \int_{-\infty}^{\infty}\log \frac{f_i(y)}{\sum_{j=1}^Iw_jf_j(y)}f_i(y)dy \\
      & = \sum_{k=1}^Kp_{ik}\int_{\kappa_k}^{\kappa_{k+1}}\log\frac{p_{ik}f_{ik}(y)}{\sum_{j=1}^{I}w_jp_{jk}f_{jk}(y)}f_{ik}(y)dy \\
      & = \sum_{k=1}^Kp_{ik}\E_{f_{ik}}\left[\log\frac{p_{ik}f_{ik}(y)}{\sum_{j=1}^{I}w_jp_{jk}f_{jk}(y)}\right] \\
      & = \sum_{k=1}^KD_{ik}.
\end{align*}
When $k$ is small enough so that each tract is uniformly distributed in bin $k$ we can solve this integral analytically. In this case we obtain \eqref{eq:app.Di.unif}
\begin{align}
  D_{ik} & = p_{ik} \int_{\kappa_k}^{\kappa_{k+1}}\log\frac{p_{ik}f_{ik}(y)}{\sum_{j=1}^{I}w_jp_{jk}f_{jk}(y)}f_{ik}(y)dy \nonumber \\
        & = p_{ik} \int_{\kappa_k}^{\kappa_{k+1}}\log\frac{\frac{p_{ik}}{\kappa_{k+1} - \kappa_{k}}}{\sum_{j=1}^Iw_j \frac{p_{jk}}{\kappa_{k+1} - \kappa_k}}\frac{1}{\kappa_{k+1} - \kappa_k} dy \nonumber\\
  & = p_{ik}\log\frac{p_{ik}}{\sum_{j=1}^Iw_jp_{jk}}. \label{eq:app.Di.unif}
\end{align}
If \emph{any} tract is not uniform distributed in a given bin, then we use \eqref{eq:app.Di.pareto} to set up the importance sampler.
\begin{align}
  D_{ik} & = p_{ik}\E_{f_{ik}}\left[\log\frac{p_{ik}f_{ik}(y)}{\sum_{j=1}^{I}w_jp_{jk}f_{jk}(y)}\right] \nonumber\\
  & = p_{ik}\E_{h_{k}}\left[\log\frac{p_{ik}f_{ik}(y)}{\sum_{j=1}^{I}w_jp_{jk}f_{jk}(y)}\frac{f_{ik}(y)}{h_k(y)}\right]. \label{eq:app.Di.pareto}
\end{align}
So, the importance weights for tract $i$ in bin $k$ are given by $f_{ik}(y) / h_k(y)$ with importance density $h_k(y)$. We choose $h_k(y)$ to be the bin $k$ density of the tract with the smallest $\alpha_{ik}$ for that bin. This ensures that the tails of the importance density dominate the tails of each tract-level density -- this is especially important in the uppermost bin.

Let $y_{kl}$ for $l=1,2,\ldots,L$ index Monte Carlo simulations from $h_k$. Then the estimator for $D_{ik}$ is given by \eqref{eq:app.Dik.estimator}.
\begin{align}
  \widehat{D}_{ik} & = p_{ik}\frac{1}{L}\sum_{l=1}^LD_{ikl}, \nonumber\\
  D_{ikl} &= \log\frac{p_{ik}f_{ik}(y_{kl})}{\sum_{j=1}^{I}w_jp_{jk}f_{jk}(y_{kl})}\frac{f_{ik}(y_{kl})}{h_k(y_{kl})}.\label{eq:app.Dik.estimator}
\end{align}
Since the same $y_{kl}$s are used for all tracts, we need to account for their error correlations. Let $\bm{C}_{D_{k}}$ denote the error covariance matrix of the $\widehat{D}_{ik}$s, with entries defined by \eqref{eq:app.Dik.cov}, where $\overline{D}_{ik} = \sum_{l=1}^LD_{ikl}/L$.
\begin{align}
\left(C_{D_k}\right)_{a,b} = \frac{1}{L}\frac{1}{L-1}\sum_{l=1}^L(D_{akl} - \overline{D}_{ak})(D_{bkl} - \overline{D}_{bk}).\label{eq:app.Dik.cov}
\end{align}
Let $k^*$ be the largest $k$ such that all tracts are uniformly distributed in bin $k$. Then our estimator for $D_i$ is given by \eqref{eq:app.Di.estimator} 
\begin{align}
\widehat{D}_i = \sum_{k=1}^{k^*}D_{ik} + \sum_{k=k^* + 1}^K\widehat{D}_{ik}. \label{eq:app.Di.estimator}
\end{align}
Again, each $\widehat{D}_i$ uses the same set of simulations, so this induces error correlation between the $\widehat{D}_i$s. Let $\bm{C}_{D}$ denote the error covariance matrix. Then it is given by \eqref{eq:app.Di.cov}
\begin{align}
\bm{C}_D = \sum_{k=k^*+1}^K\bm{C}_{D_k}.\label{eq:app.Di.cov}
\end{align}
Finally, the estimator for $D$ is given by \eqref{eq:app.D.estimator} with associated standard error given by \eqref{eq:app.D.se}, where $\bm{w} = (w_1, w_2, \dots, w_I)$
\begin{align}
  \widehat{D} & = \sum_{i=1}^Iw_i\widehat{D}_i \label{eq:app.D.estimator}\\
  S_D & = \sqrt{\bm{w}' \bm{C}_D\bm{w}}. \label{eq:app.D.se}
\end{align}

\subsection{INFORMATION THEORY INDEX}\label{app:compute.info}
First, note that
\begin{align*}
  E(F||F) &= -\int_{-\infty}^\infty F(y) \log[F(y)] + [1 - F(y)] \log[1 - F(y)] dF(y) \\
          &=-\int_{0}^1 p \log p + (1-p) \log (1-p) dp \\
          &= - \int_0^1 p\log p dp \\
          &= -2 \left[\left.\frac{1}{2}p^2 \log p \right|_0^1 - \int_0^1\frac{1}{2}p dp\right] \\
          & = 0 + \left.\frac{1}{2}p^2\right|_0^1 = \frac{1}{2}.
\end{align*}
This yields \eqref{eq:app.HR.reqrite}
\begin{align}
  H_R &= 1 - 2\sum_{i=1}^Iw_iE_i, \nonumber\\
  E_i &= -\int_{-\infty}^{\infty}\left\{F(y)\log F(y) + [1 - F(y)] \log[1 - F(y)]\right\}f_i(y)dy.  \label{eq:app.HR.reqrite}
\end{align}
To approximate these integrals, we again use importance sampling where the importance density $h(y)$ is a latent PRLN density, with $p_k = 1/K$ for $k=1,2,\dots,K$, and $\alpha_k$ set to be the smallest value of $\alpha_{ik}$ for all tracts in the metro area. If no tracts are Pareto distributed in bin $k$, then instead that bin is taken to be uniform in $h(y)$. 

Let $y_l$ for $l=1,2,\dots,L$ denote the importance sample from $L$. Then our estimator for $E_i$ is given by \eqref{eq:app.Ei.estimator}
\begin{align}
  \widehat{E}_i &= \frac{1}{L} \sum_{l=1}^LE_{il}, \nonumber\\
  E_{il} & = -\left\{F(y_l)\log F(y_l) - [1 - F(y_l)] \log[1 - F(y_l)]\right\}\frac{f_i(y_l)}{h(y_l)}.
  \label{eq:app.Ei.estimator}
\end{align}
Again, since the same importance samples were used for each tract, this induces error correlation between the $\widehat{E}_i$s. The error covariance matrix, $\bm{C}_{E}$, has entries given by \eqref{eq:app.Ei.cov}, where $\overline{E}_i = \sum_{l=1}^LE_{il} / L$ and
\begin{align}
\left(C_E\right)_{a,b} = \frac{1}{L}\frac{1}{L-1}\sum_{l=1}^L(E_{al} - \overline{E}_a)(E_{bl} - \overline{E}_b). \label{eq:app.Ei.cov}
\end{align}
Then our estimator for $H_R$ is given by \eqref{eq:app.HR.estimator} with associated standard error \eqref{eq:app.HR.se}
\begin{align}
  \widehat{H}_R &= 1 - 2\sum_{i=1}^Iw_i\widehat{E}_i \label{eq:app.HR.estimator} \\
  S_{H_R} = & 2 \sqrt{\bm{w}'\bm{C}_E\bm{w}}. \label{eq:app.HR.se}
\end{align}

\subsection{ADDITIONAL COMPUTATIONAL DETAILS}
We computed these indices for the top 100 metro areas in the U.S. by population in three distinct settings: for the household income distribution, for the black households only income distribution, and for the white hosueholds only income distribution. In each case, we used the following procedure.

For a given metro area, we fit L-PRLN to the relevant household income distribution for each Census tract within the metro area, using Stan \citep{stan2017manual} on a high performance computing cluster. We use 2000 iterations for tuning and warmup, and kept $M=2000$ iterations for inference, and obtained 4 chains in this manner. Then, we computed both $D$ and $H_R$ for all 2000 iterations of the MCMC sample. For $D$ we set $L=500$, and for $H_R$ we set $L=1000$. Standard errors for $D$ were typically about $0.3\%$ of their associated estimates, and the largest was about $1.8\%$. Standard errors for $H_R$ were typically about $2\%$ of their associated estimates, though the largest was about $24\%$. Note that these standard errors were accounted for in the EIV regressions.

These computations were parallelized in two ways. First, we fit L-PRLN and computed $D$ and $H_R$ for each chain in a separate job, so 12 jobs can be run simultaneously -- 4 chains each for all, black, and white households respectively. Second, each job had 28 cores available to it. These were used to fit the tract-level L-PRLN models in parallel, then to parallelize the computation of $D$ and $H_R$. Despite this, a single job, representing a single chain for all 100 metro areas but only one of the three possible household groups, took up to 7 days to complete. These jobs were also memory constrained because within a metro area, each Census tract's MCMC sample needs to be held in memory simultaneously to compute $D$ and $H_R$. This is particularly constraining for the New York City metro area, which contains over 4,900 Census tracts. The vast majority of the computational effort was spent computing $D$ and $H_R$, and not on fitting the L-PRLN models.

\section{SEGREGATION INDEX RESULTS}\label{app:eiv.results}
\setcounter{table}{0}

\begin{figure}[ht]
\centering
\includegraphics[scale = 0.9]{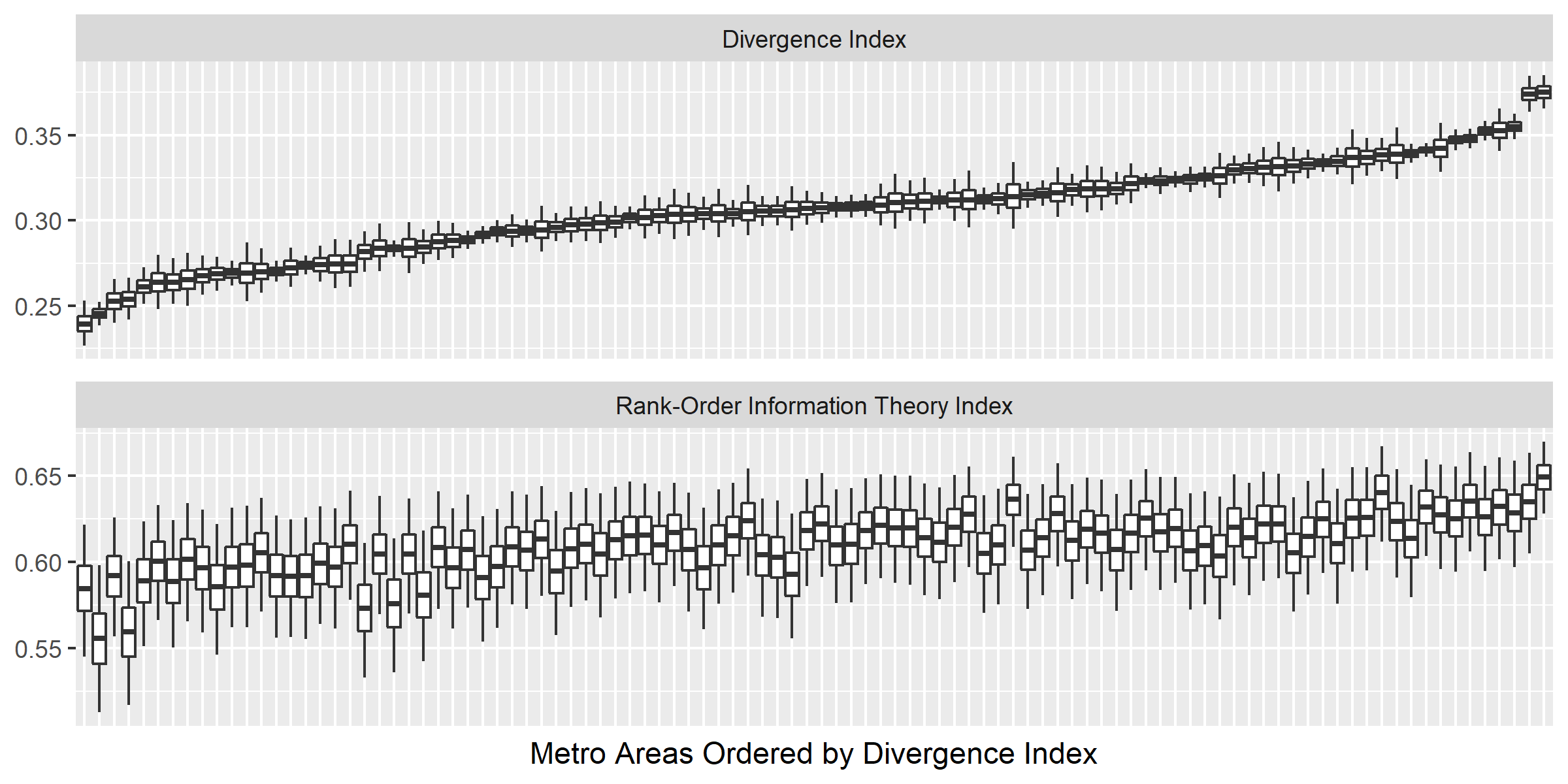}
\caption{Divergence index vs. information theory index for all households. Box plots represent the $2.5$, $25$, $50$, $75$, and $97.5$ percentiles of the posterior distribution of the index.}
\label{fig:indexplot}
\end{figure}

\begin{figure}[ht]
\centering
\includegraphics[scale = 0.9]{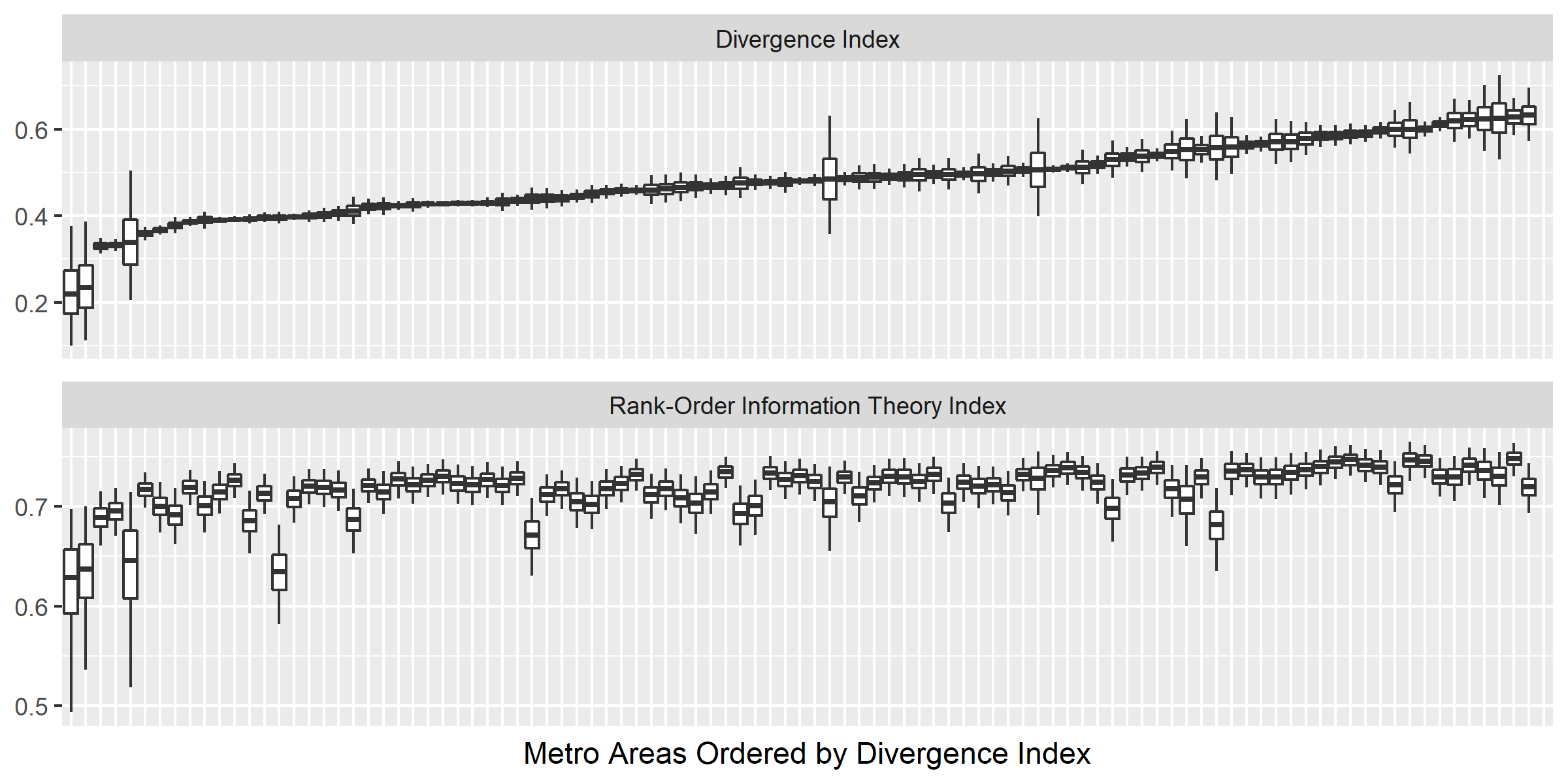}
\caption{Divergence index vs. information theory index for black households. Box plots represent the $2.5$, $25$, $50$, $75$, and $97.5$ percentiles of the posterior distribution of the index.}
\label{fig:blackindexplot}
\end{figure}

\begin{figure}[ht]
\centering
\includegraphics[scale = 0.9]{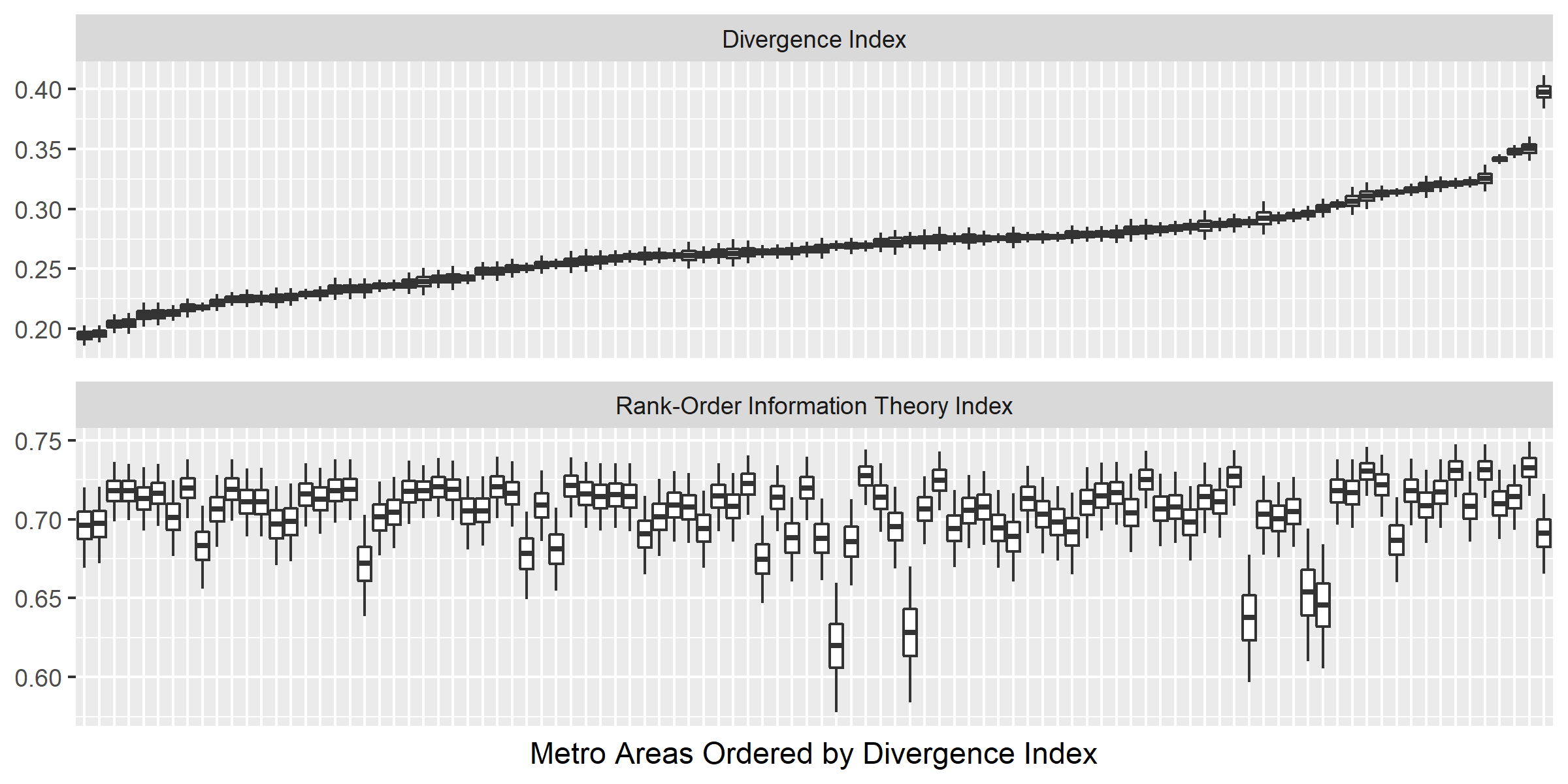}
\caption{Divergence index vs. information theory index for white households. Box plots represent the $2.5$, $25$, $50$, $75$, and $97.5$ percentiles of the posterior distribution of the index.}
\label{fig:whiteindexplot}
\end{figure}

\clearpage
\subsection{Information Theory Index Regressions}

\begin{table}[ht]
\centering
\begin{tabular}{rrrrrrrrr}
  \hline
 & OLS & Mean & SD & 2.5\% & 25\% & 50\% & 75\% & 97.5\% \\ 
  \hline
  Intercept & 0.304 & 0.237 & 0.087 & 0.068 & 0.177 & 0.236 & 0.296 & 0.406 \\ 
  Gini & 0.454 & 0.427 & 0.058 & 0.313 & 0.389 & 0.427 & 0.467 & 0.540 \\ 
  Population & 1.4E-09 & 1.2E-09 & 2.6E-10 & 7.0E-10 & 1.0E-09 & 1.2E-09 & 1.4E-09 & 1.7E-09 \\ 
  Unemp & 0.004 & 0.041 & 0.097 & -0.153 & -0.025 & 0.042 & 0.108 & 0.227 \\ 
  Edu & -0.031 & -0.010 & 0.043 & -0.095 & -0.038 & -0.010 & 0.019 & 0.073 \\ 
  Income & -1.5E-03 & -1.4E-03 & 2.6E-04 & -1.9E-03 & -1.6E-03 & -1.4E-03 & -1.2E-03 & -9.0E-04 \\ 
  AgeOver65 & -0.094 & -0.104 & 0.053 & -0.208 & -0.138 & -0.104 & -0.068 & -0.000 \\ 
  AgeUnder18 & -0.218 & -0.210 & 0.059 & -0.328 & -0.250 & -0.210 & -0.169 & -0.095 \\ 
  Foreign & -0.033 & -0.032 & 0.015 & -0.061 & -0.042 & -0.032 & -0.022 & -0.002 \\ 
  IndustyConstruct & 0.296 & 0.391 & 0.122 & 0.156 & 0.307 & 0.391 & 0.472 & 0.629 \\ 
  IndustryManuf & 0.080 & 0.085 & 0.041 & 0.005 & 0.058 & 0.085 & 0.113 & 0.165 \\ 
  IndustryFIRE & 0.028 & 0.046 & 0.046 & -0.045 & 0.016 & 0.046 & 0.077 & 0.136 \\ 
  IndustryProf & 0.077 & 0.067 & 0.036 & -0.002 & 0.043 & 0.068 & 0.091 & 0.138 \\ 
  FemaleHHer & 0.159 & 0.204 & 0.051 & 0.104 & 0.169 & 0.203 & 0.238 & 0.305 \\ 
  SameHouse & 0.087 & 0.122 & 0.067 & -0.009 & 0.077 & 0.121 & 0.168 & 0.253 \\ 
  SameCounty & 0.234 & 0.266 & 0.087 & 0.098 & 0.206 & 0.266 & 0.325 & 0.438 \\ 
  NewHouse & -0.135 & -0.122 & 0.111 & -0.337 & -0.198 & -0.124 & -0.045 & 0.096 \\ 
   \hline
\end{tabular}
\caption{OLS estimates and posterior summaries of EIV regression coefficients for the information theory index using all households.}
\label{tab:eiv.hr.raw.all}
\end{table}

\begin{table}[ht]
\centering
\begin{tabular}{rrrrrrrrr}
  \hline
 & OLS & Mean & SD & 2.5\% & 25\% & 50\% & 75\% & 97.5\% \\ 
  \hline
Intercept & 0.817 & 0.730 & 0.191 & 0.346 & 0.611 & 0.729 & 0.848 & 1.119 \\ 
  Gini & -0.191 & -0.079 & 0.125 & -0.328 & -0.161 & -0.078 & 0.004 & 0.164 \\ 
  Population & -4.6E-09 & -4.4E-09 & 3.1E-09 & -1.1E-08 & -6.5E-09 & -4.4E-09 & -2.3E-09 & 1.7E-09 \\ 
  Unemp & -0.227 & -0.203 & 0.099 & -0.398 & -0.270 & -0.203 & -0.138 & -0.009 \\ 
  Edu & 0.011 & 0.077 & 0.053 & -0.027 & 0.041 & 0.077 & 0.112 & 0.180 \\ 
  Income & 1.6E-03 & 1.4E-04 & 7.7E-03 & -1.7E-02 & -3.5E-03 & 2.1E-04 & 3.8E-03 & 1.7E-02 \\ 
  AgeOver65 & -0.066 & -0.097 & 0.129 & -0.352 & -0.184 & -0.097 & -0.009 & 0.156 \\ 
  AgeUnder18 & -0.089 & -0.005 & 0.190 & -0.381 & -0.133 & -0.005 & 0.123 & 0.367 \\ 
  Foreign & 0.045 & 0.071 & 0.025 & 0.022 & 0.055 & 0.071 & 0.088 & 0.121 \\ 
  IndustyConstruct & -0.054 & -0.189 & 0.225 & -0.628 & -0.339 & -0.190 & -0.041 & 0.263 \\ 
  IndustryManuf & -0.058 & -0.037 & 0.064 & -0.160 & -0.079 & -0.037 & 0.006 & 0.089 \\ 
  IndustryFIRE & 0.142 & 0.144 & 0.088 & -0.032 & 0.085 & 0.144 & 0.203 & 0.318 \\ 
  IndustryProf & -0.063 & -0.037 & 0.052 & -0.138 & -0.073 & -0.038 & -0.003 & 0.066 \\ 
  SameHouse & 0.053 & 0.038 & 0.127 & -0.214 & -0.046 & 0.038 & 0.123 & 0.288 \\ 
  SameCounty & -0.068 & -0.255 & 0.194 & -0.632 & -0.384 & -0.256 & -0.124 & 0.124 \\ 
  NewHouse & 0.021 & 0.259 & 0.243 & -0.223 & 0.096 & 0.263 & 0.424 & 0.728 \\ 
   \hline
\end{tabular}
\caption{OLS estimates and posterior summaries of EIV regression coefficients for the information theory index using only black households.}
\label{tab:eiv.hr.raw.black}
\end{table}

\begin{table}[ht]
\centering
\begin{tabular}{rrrrrrrrr}
  \hline
 & OLS & Mean & SD & 2.5\% & 25\% & 50\% & 75\% & 97.5\% \\ 
  \hline
Intercept & 0.682 & 0.955 & 0.230 & 0.491 & 0.813 & 0.956 & 1.098 & 1.415 \\ 
  Gini & 0.370 & 0.145 & 0.111 & -0.072 & 0.071 & 0.146 & 0.220 & 0.366 \\ 
  Population & 1.6E-09 & 2.0E-09 & 9.6E-10 & 1.3E-10 & 1.4E-09 & 2.0E-09 & 2.7E-09 & 3.9E-09 \\ 
  Unemp & -0.203 & 0.157 & 0.231 & -0.296 & 0.003 & 0.156 & 0.311 & 0.611 \\ 
  Edu & -0.021 & -0.243 & 0.060 & -0.361 & -0.283 & -0.242 & -0.202 & -0.124 \\ 
  Income & -2.5E-03 & 7.0E-05 & 5.5E-03 & -1.1E-02 & -3.0E-03 & -8.9E-05 & 3.1E-03 & 1.2E-02 \\ 
  AgeOver65 & -0.152 & 0.008 & 0.073 & -0.137 & -0.041 & 0.008 & 0.056 & 0.153 \\ 
  AgeUnder18 & -0.387 & -0.254 & 0.094 & -0.438 & -0.317 & -0.255 & -0.193 & -0.069 \\ 
  Foreign & -0.057 & -0.154 & 0.026 & -0.205 & -0.171 & -0.154 & -0.137 & -0.104 \\ 
  IndustyConstruct & 0.178 & -0.226 & 0.212 & -0.641 & -0.369 & -0.226 & -0.085 & 0.190 \\ 
  IndustryManuf & 0.004 & -0.177 & 0.063 & -0.299 & -0.220 & -0.178 & -0.136 & -0.054 \\ 
  IndustryFIRE & 0.083 & -0.008 & 0.080 & -0.166 & -0.062 & -0.008 & 0.045 & 0.149 \\ 
  IndustryProf & -0.017 & -0.215 & 0.055 & -0.321 & -0.252 & -0.215 & -0.179 & -0.106 \\ 
  SameHouse & 0.058 & 0.060 & 0.129 & -0.194 & -0.026 & 0.060 & 0.146 & 0.312 \\ 
  SameCounty & 0.246 & 0.557 & 0.174 & 0.218 & 0.441 & 0.558 & 0.673 & 0.897 \\ 
  NewHouse & -0.151 & 0.044 & 0.199 & -0.347 & -0.088 & 0.044 & 0.177 & 0.434 \\ 
   \hline
\end{tabular}
\caption{OLS estimates and posterior summaries of EIV regression coefficients for the information index using only white households.}
\label{tab:eiv.hr.raw.white}
\end{table}

\clearpage

\subsection{Divergence Index Regressions}

\begin{table}[ht]
\centering
\begin{tabular}{rrrrrrrrr}
  \hline
 & OLS & Mean & SD & 2.5\% & 25\% & 50\% & 75\% & 97.5\% \\ 
  \hline
Intercept & -0.123 & -0.149 & 0.230 & -0.609 & -0.299 & -0.150 & -0.000 & 0.315 \\ 
  Gini & 0.765 & 0.763 & 0.157 & 0.458 & 0.657 & 0.762 & 0.868 & 1.074 \\ 
  Population & 2.9E-09 & 2.9E-09 & 9.0E-10 & 1.1E-09 & 2.3E-09 & 2.9E-09 & 3.5E-09 & 4.6E-09 \\ 
  Unemp & 0.508 & 0.496 & 0.292 & -0.067 & 0.298 & 0.496 & 0.688 & 1.068 \\ 
  Edu & -0.009 & -0.025 & 0.114 & -0.253 & -0.101 & -0.025 & 0.052 & 0.197 \\ 
  Income & -7.1E-04 & -6.8E-04 & 7.7E-04 & -2.2E-03 & -1.2E-03 & -6.8E-04 & -1.6E-04 & 8.2E-04 \\ 
  AgeOver65 & -0.243 & -0.232 & 0.131 & -0.484 & -0.321 & -0.232 & -0.145 & 0.023 \\ 
  AgeUnder18 & 0.048 & 0.053 & 0.153 & -0.252 & -0.050 & 0.053 & 0.157 & 0.351 \\ 
  Foreign & -0.088 & -0.090 & 0.051 & -0.191 & -0.125 & -0.091 & -0.056 & 0.009 \\ 
  IndustyConstruct & 0.341 & 0.410 & 0.330 & -0.232 & 0.194 & 0.404 & 0.629 & 1.061 \\ 
  IndustryManuf & 0.074 & 0.084 & 0.108 & -0.125 & 0.011 & 0.082 & 0.157 & 0.296 \\ 
  IndustryFIRE & 0.063 & 0.066 & 0.120 & -0.171 & -0.013 & 0.064 & 0.147 & 0.304 \\ 
  IndustryProf & 0.050 & 0.062 & 0.095 & -0.124 & -0.000 & 0.063 & 0.125 & 0.246 \\ 
  FemaleHHer & 0.122 & 0.152 & 0.133 & -0.119 & 0.065 & 0.151 & 0.240 & 0.406 \\ 
  SameHouse & -0.049 & -0.038 & 0.188 & -0.407 & -0.163 & -0.036 & 0.086 & 0.336 \\ 
  SameCounty & 0.464 & 0.512 & 0.259 & -0.012 & 0.341 & 0.513 & 0.685 & 1.020 \\ 
  NewHouse & -0.584 & -0.581 & 0.305 & -1.185 & -0.785 & -0.579 & -0.376 & 0.015 \\ 
   \hline
\end{tabular}
\caption{OLS estimates and posterior summaries of EIV regression coefficients for the divergence index using all households.}
\label{tab:eiv.kl.raw.all}
\end{table}

\begin{table}[ht]
\centering
\begin{tabular}{rrrrrrrrr}
  \hline
 & OLS & Mean & SD & 2.5\% & 25\% & 50\% & 75\% & 97.5\% \\ 
  \hline
Intercept & -0.297 & -0.416 & 1.033 & -2.484 & -1.082 & -0.409 & 0.254 & 1.618 \\ 
  Gini & 0.699 & 1.403 & 0.742 & -0.036 & 0.899 & 1.402 & 1.898 & 2.872 \\ 
  Population & -6.2E-08 & -5.8E-08 & 1.6E-08 & -9.0E-08 & -6.9E-08 & -5.8E-08 & -4.8E-08 & -2.7E-08 \\ 
  Unemp & -1.156 & -1.545 & 0.531 & -2.584 & -1.902 & -1.546 & -1.195 & -0.485 \\ 
  Edu & 0.564 & 0.387 & 0.256 & -0.118 & 0.216 & 0.388 & 0.559 & 0.894 \\ 
  Income & -4.3E-03 & -2.6E-03 & 4.2E-02 & -8.8E-02 & -2.8E-02 & -4.5E-03 & 2.3E-02 & 8.6E-02 \\ 
  AgeOver65 & 0.037 & 0.214 & 0.508 & -0.773 & -0.128 & 0.210 & 0.551 & 1.220 \\ 
  AgeUnder18 & -0.153 & -0.009 & 0.809 & -1.602 & -0.550 & -0.014 & 0.530 & 1.596 \\ 
  Foreign & 0.459 & 0.402 & 0.118 & 0.166 & 0.323 & 0.403 & 0.480 & 0.633 \\ 
  IndustyConstruct & -0.046 & -0.475 & 1.131 & -2.711 & -1.233 & -0.473 & 0.288 & 1.722 \\ 
  IndustryManuf & 0.313 & 0.195 & 0.315 & -0.424 & -0.016 & 0.197 & 0.406 & 0.819 \\ 
  IndustryFIRE & 0.330 & 0.380 & 0.402 & -0.404 & 0.107 & 0.381 & 0.649 & 1.168 \\ 
  IndustryProf & 0.228 & 0.044 & 0.271 & -0.483 & -0.136 & 0.043 & 0.225 & 0.578 \\ 
  SameHouse & -0.049 & -0.084 & 0.647 & -1.346 & -0.519 & -0.086 & 0.348 & 1.198 \\ 
  SameCounty & 0.079 & 0.258 & 0.944 & -1.595 & -0.371 & 0.254 & 0.889 & 2.122 \\ 
  NewHouse & 0.533 & 0.556 & 1.098 & -1.619 & -0.169 & 0.557 & 1.286 & 2.723 \\ 
   \hline
\end{tabular}
\caption{OLS estimates and posterior summaries of EIV regression coefficients for the divergence index using black households only.}
\label{tab:eiv.kl.raw.black}
\end{table}

\begin{table}[ht]
\centering
\begin{tabular}{rrrrrrrrr}
  \hline
 & OLS & Mean & SD & 2.5\% & 25\% & 50\% & 75\% & 97.5\% \\ 
  \hline
Intercept & 0.405 & 0.188 & 0.440 & -0.660 & -0.101 & 0.179 & 0.471 & 1.079 \\ 
  Gini & 0.489 & 0.683 & 0.240 & 0.207 & 0.523 & 0.683 & 0.843 & 1.155 \\ 
  Population & 4.5E-09 & 4.3E-09 & 2.1E-09 & 1.4E-10 & 2.9E-09 & 4.3E-09 & 5.7E-09 & 8.4E-09 \\ 
  Unemp & 0.473 & 0.356 & 0.497 & -0.630 & 0.024 & 0.356 & 0.690 & 1.326 \\ 
  Edu & -0.098 & -0.001 & 0.127 & -0.251 & -0.087 & -0.001 & 0.084 & 0.247 \\ 
  Income & 1.2E-03 & 1.4E-03 & 9.6E-03 & -1.9E-02 & -4.4E-03 & 2.5E-03 & 7.1E-03 & 2.0E-02 \\ 
  AgeOver65 & -0.567 & -0.638 & 0.154 & -0.939 & -0.741 & -0.638 & -0.535 & -0.338 \\ 
  AgeUnder18 & -0.091 & -0.146 & 0.197 & -0.531 & -0.279 & -0.146 & -0.014 & 0.239 \\ 
  Foreign & -0.039 & 0.002 & 0.055 & -0.107 & -0.035 & 0.002 & 0.039 & 0.112 \\ 
  IndustyConstruct & 0.286 & 0.530 & 0.458 & -0.374 & 0.225 & 0.531 & 0.835 & 1.429 \\ 
  IndustryManuf & -0.205 & -0.104 & 0.134 & -0.366 & -0.195 & -0.104 & -0.014 & 0.159 \\ 
  IndustryFIRE & -0.071 & -0.020 & 0.168 & -0.350 & -0.131 & -0.020 & 0.093 & 0.307 \\ 
  IndustryProf & -0.158 & -0.049 & 0.118 & -0.281 & -0.128 & -0.049 & 0.029 & 0.182 \\ 
  SameHouse & -0.200 & -0.215 & 0.277 & -0.760 & -0.400 & -0.217 & -0.031 & 0.327 \\ 
  SameCounty & 0.077 & -0.067 & 0.377 & -0.806 & -0.320 & -0.069 & 0.185 & 0.675 \\ 
  NewHouse & -0.381 & -0.492 & 0.432 & -1.347 & -0.775 & -0.491 & -0.207 & 0.352 \\ 
   \hline
\end{tabular}
\caption{OLS estimates and posterior summaries of EIV regression coefficients for the divergence index using white households only.}
\label{tab:eiv.kl.raw.white}
\end{table}

\clearpage

\bibliographystyle{jasa}  
\bibliography{lprln}

\end{document}